# Multiscale energy transfers in a forced two-DOF oscillatory system with a grounded nonlinear energy sink

Lan Huang [a, *]     Xiaodong Yang [b, c]

a Department of Mechanics, Inner Mongolia University of Technology, Hohhot 010051, PR China
b Beijing Key Laboratory of Nonlinear Vibrations and Strength of Mechanical Structures, Department of Mechanics, Beijing University of Technology, Beijing 100124, PR China
c College of Aerospace Engineering, Shenyang Aerospace University, Shenyang 110136, PR China

**Abstract:** This paper studies the steady-state oscillations and the corresponding energy transfers in the two-DOF mechanical system with a grounded nonlinear energy sink (NES), under an external excitation. Based on the complexification-multiscale method, the slow frequency and the fast frequency of the complex amplitudes on the invariant manifold are analyzed, without omitting any terms. The analytical results show that the complex amplitudes of the primary system responses contain only the slow frequency, while that of the NES responses can contain both the slow frequency and the fast frequency when it triggers the high-order frequency term. In addition, each type of set on the invariant manifold represents the corresponding steady-state oscillations of the system, that is, the fixed point corresponds to the stationary oscillation, the interval corresponds to the nonlinear beat, and the hysteresis loop corresponds to the relaxation oscillation. Accordingly, we rewrite the forms of the complexification system to match the corresponding types of steady-state oscillations. Furthermore, the Hilbert-Huang transform is used to investigate the time-frequency relation of the displacements and the energies of the three steady-state responses. The Hilbert spectrums indicate that there is no energy transfer in the stationary oscillations, the $O(\varepsilon)/O(\varepsilon)$ type of energy transfer in the nonlinear beats, and the $O(\varepsilon)/O(\varepsilon)$ and the $O(\varepsilon)/O(1)$ types of energy transfer in the relaxation oscillations. The theoretical results are confirmed by numerical simulations.

E-mail address:
lanhuang0919@163.com ( L. Huang, ∗ Corresponding author)
jxdyang@163.com ( X.-D. Yang)

# 1. Introduction

Considerable attention has been given to the use of a nonlinear energy sink (NES) [1] as a nonlinear vibration absorber. An NES is an efficient passive control device, which has been the focus of extensive research [2]. Unlike the linear vibration absorber, the NESs are dynamic vibration absorbers that achieve vibration suppression for a wide range of frequencies and energy. The efficiency of NES systems is verified experimentally as well in engineering [3-5]. According to the type of nonlinear coupling mechanism with the primary, the NESs can be classified: the stiffness-based NESs [6], rotary NESs [7], impact-based NESs [8], and other kinds of NESs [9-13].

For investigating the dynamics and the passive vibration energy transfer [14] of the systems with NESs, various analytical and numerical methods have been applied, including the direct time integration method [15], harmonic balance method [16], complex-averaging method [1], and the multiscale method [17]. Especially, the combined application of the complex variables and the multiscale method is efficient in studying the multiscale dynamics and vibration reduction mechanisms of the mechanical systems containing NESs. For instance, using the complexification-multiscale method, the multiscale dynamics of a non-smooth master system which is coupled to a non-smooth NES during free and forced oscillations is studied analytically and numerically by Lamarque et al [18]. Multi-scale energy exchanges between an elasto-plastic oscillator and a light non-smooth system with external pre-stress have been proposed by Weiss et al [19], combining the multiscale method and the time event-driven technique. In [20], the method, combining the complex-averaging and asymptotic analysis of the slow-flow-based geometric singular perturbation theory, is first applied to the prediction of the mitigation limit of a system subject to friction-induced vibrations coupled to two NES.

In the monograph [21], Manevich et al. discussed in detail the complexification-multiscale method and its application in the analytical investigation of the coupled steady-state modes of vibratory systems with internal resonance. For the two-DOF symmetric cubic system in 1:1 internal resonance, the dynamics of the system on the zero-order time (i.e., fast time $\tau_0$) will not occur since the derivatives of both the complex amplitudes with respect to the fast time are equal to 0. This symmetry causes the complex amplitudes to perform the dynamics on the same time scale, e.g., $\varphi_{i0}(\tau_0)=0$ and $\varphi_{i0}(\tau_1)\neq 0$ ($i=1,2$). It is completely different from the systems with NESs, because the ratio of the masses between the NES and the primary system is very small, which leads to the occurrence of high asymmetry in the systems. For instance, in a two-DOF system with an NES whose mass is far less than that of the primary system, by using the complexification-multiscale method, the derivative of the complex amplitude of the primary system with respect to the fast time is equal to 0, i.e., $\varphi_{10}(\tau_0)=0$, while that of the complex amplitude of NES is not 0, i.e., $\varphi_{20}(\tau_0)\neq 0$. In addition, both of the derivatives of the two complex amplitudes with respect to the slow time are not equal to 0, i.e., $\varphi_{10}(\tau_1)\neq 0$, $\varphi_{20}(\tau_1)\neq 0$. This implies that energy transfers on different time scales may occur in the system.

Recently, Huang et al. [12] studied a two-DOF mechanical system with a grounded NES, under a harmonically external forcing in the regime of 1:1:1 resonance. Based on the complexification-averaging method and the bifurcation analysis, the qualitative results of several bifurcation boundaries of the slow flow are detected. They further analyzed the dynamics of the slow flow on different time scales and discussed the necessary condition of occurrence of the strongly modulated responses. The two singular points of the slow flow on the slow time have been verified to be a pair of SN bifurcation points essentially. Furthermore, to investigate the multiscale energy transfer mechanism during the system responses, e.g., the nonlinear beats and the strongly modulated responses, Huang et al. [22] studied the modal interactions and the energy transfers between the primary system and the NES in this system. Using the fast-slow analysis, the slow flow has been rewritten as the fast-slow systems, defined on different time scales, to obtain the critical manifold that can capture the multiscale dynamics in the system. To go further, the numerical results and the Hilbert spectrums have validated theoretical predictions that the system undergoes multi-scale energy transfers. Finally, three types of energy transfer are defined, namely three-time-scale energy transfer, two-time-scale energy transfer, and single-time-scale energy transfer. In addition, two energy transfer modes corresponding to special oscillations were discussed, namely no energy transfer and energy exchange.

However, in the article [22], the mechanism of energy transfers on different time scales has not been fully explored. Specifically, the primary system and the NES can perform vastly different types of oscillations. The complex amplitude of the primary system is always slowly varying, while that of the NES can either only perform slow oscillations or simultaneously perform both slow and fast oscillations. This means that the complex amplitudes of the primary system and the NES may contain different frequencies that play a core role in the mechanism for the multiscale energy transfer. This has prompted us to further explore.

In this work, thus, we further study the two-DOF system with a grounded NES to investigate the steady-state oscillations in the system and to discuss the mechanism of multiscale energy transfers in these different types of oscillations.

The equations of motion of the system are expressed as:

$$\begin{cases} m_1\ddot{x}_1 + k_1'x_1 + c_1'\dot{x}_1 + k_2'(x_1 - x_2) + c_2'(\dot{x}_1 - \dot{x}_2) = f\cos(\Omega t) \\ m_2\ddot{x}_2 + k_2'(x_2 - x_1) + c_2'(\dot{x}_2 - \dot{x}_1) + k'x_2^3 + c'x_2^2\dot{x}_2 = 0 \end{cases} \quad (1)$$

In Eq.(1), the mass $m_2$ is far less than the mass $m_1$ of the primary system, i.e., $m_2/m_1=\varepsilon$ ($0<\varepsilon\ll 1$). Physically, it opens the possibility that the energy is transferred from the large mass to the small mass in an asymmetric manner. The dimensionless form of Eq.(1) is shown as:

$$\begin{cases} \ddot{x}_1 + x_1 + \varepsilon\lambda_1\dot{x}_1 + \varepsilon k_2(x_1 - x_2) + \varepsilon\lambda_2(\dot{x}_1 - \dot{x}_2) = \varepsilon A\cos\omega\tau \\ \varepsilon\ddot{x}_2 + \varepsilon k_2(x_2 - x_1) + \varepsilon\lambda_2(\dot{x}_2 - \dot{x}_1) + \varepsilon k x_2^3 + \varepsilon\lambda x_2^2\dot{x}_2 = 0 \end{cases} \quad (2)$$

In Eq.(2), under the external excitation whose frequency ($\omega=1+\varepsilon\sigma$) is close to the natural frequency ($\omega_n=1$) of the primary system, the primary system and the NES both perform the responses that vibrate at the natural frequency. Combining the asymmetric manner of energy transfer in this system, we find that the amplitude of the primary system response and that of the NES response may change in vastly distinct time scales.

Therefore, in the paper, using the complex variables and the multiscale method, we study the system without omitting any terms, compared to the complexification-averaging method. Furthermore, the invariant manifold is introduced to capture the steady-state oscillations of the system. By investigating the invariant manifold and the high-order frequency term, we find that there are several types of steady-state oscillations, namely stationary oscillations, nonlinear beats, and relaxation oscillations, in the system responses. The first two types of these oscillations did not trigger the high-order frequency term, compared to the relaxation oscillations where only the NES presents the relaxation oscillations, while the primary system presents the nonlinear beats. Accordingly, we rewrite the form of the complexification system to match these three types of steady-state oscillations.

Furthermore, we use the Hilbert-Huang transform (HHT) [23] to explore the time-frequency relation in the energy transfers of these steady-state oscillations. Similar to frequency-energy plots (FEPs) [14] obtained by wavelet transform (WT), HHT is also a good choice for analyzing the modal interactions and the energy transfers in strongly nonlinear systems, under external excitation. Note that Eq.(2) can be seen as a two-DOF system consisting of a linear oscillator that is coupled with a strongly nonlinear oscillator [22]. HHT is a superior tool for time-frequency analysis of nonlinear and nonstationary responses because it is based on an adaptive basis. Consequently, the Hilbert spectrums, corresponding to the displacements and the energies of the three types of steady-state oscillations, indicate that there is no energy transfer in the stationary oscillations. Differently, the nonlinear beats perform a weak energy transfer, and the relaxation oscillations perform an intensive energy transfer.

The organization of the paper is as follows: in Section 2, we describe the model of the system under consideration, with the dimensionless and the complexification forms. The multiple time scale analysis of the complexification system is studied in Section 3. In Section 4, three types of steady-state oscillations in the system are discussed and the corresponding forms of the complexification system are rewritten, respectively. Detections of different types of energy transfers in the steady-state oscillations by the Hilbert spectrums are summarized in Section 5. Finally, conclusions are collected in Section 6.

## 2. Model

Consider a two-DOF mechanical system consisting of a primary system and a grounded NES as a vibration absorber. (see Fig. 1).

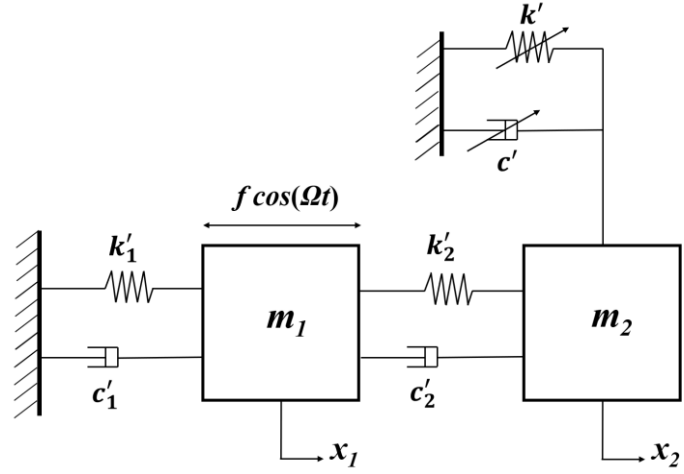

Fig.1. The two-DOF system consists of a primary system under an external excitation coupled to a grounded NES.

The equations of motion of the system are presented as:

$$\begin{cases} m_1\ddot{x}_1 + k'_1 x_1 + c'_1 \dot{x}_1 + k'_2(x_1 - x_2) + c'_2(\dot{x}_1 - \dot{x}_2) = f\cos(\Omega t) \\ m_2\ddot{x}_2 + k'_2(x_2 - x_1) + c'_2(\dot{x}_2 - \dot{x}_1) + k'x_2^3 + c'x_2^2\dot{x}_2 = 0 \end{cases} \quad (3)$$

$m_1$ and $m_2$ are the masses for the primary system and the NES, respectively. $x_1$ is the displacement of the primary system. $x_2$ is the displacement of the NES. $c'$, $c'_1$, and $c'_2$ are the damping. $k'$ and $k'_2$ are the stiffness. $f\cos(\Omega t)$ stands for the harmonically external forcing.

$$\begin{cases} \omega_0^2 = \dfrac{k'_1}{m_1}, \omega_0 t = \tau, \dfrac{c'_1}{m_1\omega_0} = \varepsilon\lambda_1, \dfrac{k'_2}{m_1\omega_0^2} = \varepsilon k_2, \dfrac{c'_2}{m_1\omega_0} = \varepsilon\lambda_2, \\ \dfrac{k'}{m_1\omega_0^2} = \varepsilon k, \dfrac{c'}{m_1\omega_0} = \varepsilon\lambda, \dfrac{f}{m_1\omega_0^2} = \varepsilon A, \omega = \dfrac{\Omega}{\omega_0} = 1 + \varepsilon\sigma, 0 < \varepsilon \ll 1. \end{cases} \quad (4)$$

By using Eq.(4), the non-dimensional equations of motion of Eq.(3) are rewritten as:

$$\begin{cases} \ddot{x}_1 + x_1 + \varepsilon\lambda_1\dot{x}_1 + \varepsilon k_2(x_1 - x_2) + \varepsilon\lambda_2(\dot{x}_1 - \dot{x}_2) = \varepsilon A\cos\omega\tau \\ \varepsilon\ddot{x}_2 + \varepsilon k_2(x_2 - x_1) + \varepsilon\lambda_2(\dot{x}_2 - \dot{x}_1) + \varepsilon k x_2^3 + \varepsilon\lambda x_2^2\dot{x}_2 = 0 \end{cases} \quad (5)$$

The complex variables [21] are introduced as follows:

$$\begin{cases} \psi_i = \dot{x}_i + j\omega x_i & j = \sqrt{-1}, i = 1,2 \\ \psi_i^* = \dot{x}_i - j\omega x_i & j = \sqrt{-1}, i = 1,2 \end{cases} \quad (6)$$

Substituting Eq.(6) into Eq.(5) we get the complexification system:

$$\begin{cases} \dot{\psi}_1 + \dfrac{\varepsilon\lambda_1 - j\omega}{2}(\psi_1 + \psi_1^*) + \dfrac{1}{2j\omega}(\psi_1 - \psi_1^*) + \dfrac{\varepsilon k_2}{2j\omega}(\psi_1 - \psi_1^* - \psi_2 + \psi_2^*) + \dfrac{\varepsilon\lambda_2}{2}(\psi_1 + \psi_1^* - \psi_2 - \psi_2^*) \\ \quad = \dfrac{\varepsilon A}{2}(e^{j\omega\tau} + e^{-j\omega\tau}) \\ \varepsilon\dot{\psi}_2 - \dfrac{j\varepsilon\omega}{2}(\psi_2 + \psi_2^*) - \dfrac{\varepsilon k_2}{2j\omega}(\psi_1 - \psi_1^* - \psi_2 + \psi_2^*) - \dfrac{\varepsilon\lambda_2}{2}(\psi_1 + \psi_1^* - \psi_2 - \psi_2^*) + \dfrac{\varepsilon k}{(2j\omega)^3}(\psi_2 - \psi_2^*)^3 \\ \quad + \dfrac{\varepsilon\lambda}{2(2j\omega)^2}(\psi_2 - \psi_2^*)^2(\psi_2 + \psi_2^*) = 0 \end{cases} \quad (7)$$

Putting

$$\begin{cases} \psi_i = \varphi_i e^{j\omega\tau} & i = 1,2 \\ \psi_i^* = \varphi_i^* e^{-j\omega\tau} & i = 1,2 \end{cases} \quad (8)$$

we come to the set of equations:

$$\begin{cases} \dot{\varphi}_1 + j\omega\varphi_1 + \dfrac{\varepsilon\lambda_1 - j\omega}{2}(\varphi_1 + \varphi_1^* e^{-2j\omega\tau}) + \dfrac{1}{2j\omega}(\varphi_1 - \varphi_1^* e^{-2j\omega\tau}) + \dfrac{\varepsilon k_2}{2j\omega}(\varphi_1 - \varphi_1^* e^{-2j\omega\tau} - \varphi_2 + \varphi_2^* e^{-2j\omega\tau}) \\ \quad + \dfrac{\varepsilon\lambda_2}{2}(\varphi_1 + \varphi_1^* e^{-2j\omega\tau} - \varphi_2 - \varphi_2^* e^{-2j\omega\tau}) = \dfrac{\varepsilon A}{2}(1 + e^{-2j\omega\tau}) \\ \varepsilon\dot{\varphi}_2 + j\varepsilon\omega\varphi_2 - \dfrac{j\varepsilon\omega}{2}(\varphi_2 + \varphi_2^* e^{-2j\omega\tau}) - \dfrac{\varepsilon k_2}{2j\omega}(\varphi_1 - \varphi_1^* e^{-2j\omega\tau} - \varphi_2 + \varphi_2^* e^{-2j\omega\tau}) - \dfrac{\varepsilon\lambda_2}{2}(\varphi_1 + \varphi_1^* e^{-2j\omega\tau} - \varphi_2 \\ \quad - \varphi_2^* e^{-2j\omega\tau}) + \dfrac{\varepsilon k}{(2j\omega)^3}(\varphi_2^3 e^{2j\omega\tau} - 3\varphi_2^2\varphi_2^* + 3\varphi_2\varphi_2^{*2} e^{-2j\omega\tau} - \varphi_2^{*3} e^{-4j\omega\tau}) + \dfrac{\varepsilon\lambda}{2(2j\omega)^2}(\varphi_2^3 e^{2j\omega\tau} - \varphi_2^2\varphi_2^* \\ \quad - \varphi_2\varphi_2^{*2} e^{-2j\omega\tau} + \varphi_2^{*3} e^{-4j\omega\tau}) = 0 \end{cases} \quad (9)$$

## 3. Multiple time scale analysis for the complexification system

### 3.1. Multiple time scales technique

Introducing the multi-time scales $\tau_i = \varepsilon^i \tau$ ($i = 0,1,2\ldots$), and taking into account the expansion of the differential operator:

$$\frac{d}{d\tau} = \frac{d}{d\tau_0} + \varepsilon \frac{d}{d\tau_1} + \ldots \quad (10)$$

we seek the solution in the form:

$$\varphi_l = \varphi_{l0}(\tau_0, \tau_1, \cdots) + \varepsilon\varphi_{l1}(\tau_0, \tau_1, \cdots) + \cdots \quad (l = 1,2) \quad (11)$$

By substituting Eq.(10) and Eq.(11) into Eq.(9), then, we obtain:

$$\tau_0: \begin{cases} \dfrac{d\varphi_{10}}{d\tau_0} = 0 \\ \dfrac{d\varphi_{20}}{d\tau_0} + \dfrac{1}{2}j\varphi_{20} - (\dfrac{\lambda_2}{2} - j\dfrac{k_2}{2\omega})(\varphi_{10} - \varphi_{20}) + (\dfrac{\lambda}{8\omega^2} - j\dfrac{3k}{8\omega^3})\varphi_{20}^2\varphi_{20}^* - \dfrac{1}{2}j\varphi_{20}^* e^{-2j\omega\tau_0} \\ \quad - (\dfrac{\lambda_2}{2} + j\dfrac{k_2}{2\omega})(\varphi_{10}^* - \varphi_{20}^*)e^{-2j\omega\tau_0} - (\dfrac{\lambda}{8\omega^2} - j\dfrac{k}{8\omega^3})\varphi_{20}^3 e^{2j\omega\tau_0} + (\dfrac{\lambda}{8\omega^2} + j\dfrac{3k}{8\omega^3})\varphi_{20}\varphi_{20}^{*2}e^{-2j\omega\tau_0} \\ \quad - (\dfrac{\lambda}{8\omega^2} + j\dfrac{k}{8\omega^3})\varphi_{20}^{*3}e^{-4j\omega\tau_0} = 0 \end{cases} \quad (12)$$

and

$$\tau_1: \begin{cases} \dfrac{d\varphi_{11}}{d\tau_0} + \dfrac{d\varphi_{10}}{d\tau_1} + (\dfrac{\lambda_1}{2} + j\dfrac{\sigma}{\omega})\varphi_{10} + (\dfrac{\lambda_1}{2} - j\dfrac{\sigma}{\omega})\varphi_{10}^* e^{-2j\omega\tau_0} + (\dfrac{\lambda_2}{2} - j\dfrac{k_2}{2\omega})(\varphi_{10} - \varphi_{20}) \\ \quad + (\dfrac{\lambda_2}{2} + j\dfrac{k_2}{2\omega})(\varphi_{10}^* e^{-2j\omega\tau_0} - \varphi_{20}^* e^{-2j\omega\tau_0}) = \dfrac{A}{2}(1 + e^{-2j\omega\tau_0}) \\ \dfrac{d\varphi_{21}}{d\tau_0} + \dfrac{d\varphi_{20}}{d\tau_1} + \dfrac{1}{2}j\sigma\varphi_{20} - \dfrac{1}{2}j\sigma\varphi_{20}^* e^{-2j\omega\tau_0} = 0 \end{cases} \quad (13)$$

The conditions [21] of the absence of secular terms in Eq.(13) give the set of equations:

$$\begin{cases} \dfrac{d\varphi_{10}}{d\tau_1} + (\dfrac{\lambda_1}{2} + j\dfrac{\sigma}{\omega})\varphi_{10} + (\dfrac{\lambda_2}{2} - j\dfrac{k_2}{2\omega})(\varphi_{10} - \varphi_{20}) = \dfrac{A}{2} \\ \dfrac{d\varphi_{20}}{d\tau_1} + \dfrac{1}{2}j\sigma\varphi_{20} = 0 \end{cases} \quad (14)$$

Furthermore, we get the asymptotic dynamics, denoted on the fast time $\tau_0$ and the slow time $\tau_1$, of zero order of complex amplitudes $\varphi_{10}$ and $\varphi_{20}$, respectively. They are shown as:

$$\tau_0: \begin{cases} \dfrac{d\varphi_{10}}{d\tau_0} = 0 \\ \dfrac{d\varphi_{20}}{d\tau_0} + \dfrac{1}{2}j\varphi_{20} - (\dfrac{\lambda_2}{2} - j\dfrac{k_2}{2\omega})(\varphi_{10} - \varphi_{20}) + (\dfrac{\lambda}{8\omega^2} - j\dfrac{3k}{8\omega^3})\varphi_{20}^2\varphi_{20}^* - \dfrac{1}{2}j\varphi_{20}^* e^{-2j\omega\tau_0} \\ \quad - (\dfrac{\lambda_2}{2} + j\dfrac{k_2}{2\omega})(\varphi_{10}^* - \varphi_{20}^*)e^{-2j\omega\tau_0} - (\dfrac{\lambda}{8\omega^2} - j\dfrac{k}{8\omega^3})\varphi_{20}^3 e^{2j\omega\tau_0} + (\dfrac{\lambda}{8\omega^2} + j\dfrac{3k}{8\omega^3})\varphi_{20}\varphi_{20}^{*2}e^{-2j\omega\tau_0} \\ \quad - (\dfrac{\lambda}{8\omega^2} + j\dfrac{k}{8\omega^3})\varphi_{20}^{*3}e^{-4j\omega\tau_0} = 0 \end{cases} \quad (15)$$

and

$$\tau_1: \begin{cases} \dfrac{d\varphi_{10}}{d\tau_1} + (\dfrac{\lambda_1}{2} + j\dfrac{\sigma}{\omega})\varphi_{10} + (\dfrac{\lambda_2}{2} - j\dfrac{k_2}{2\omega})(\varphi_{10} - \varphi_{20}) = \dfrac{A}{2} \\ \dfrac{d\varphi_{20}}{d\tau_1} + \dfrac{1}{2}j\sigma\varphi_{20} = 0 \end{cases} \quad (16)$$

In Eq.(15), the derivative of $\varphi_{10}$ with respect to $\tau_0$ is equal to zero, implying there are no dynamics for $\varphi_{10}$ on the fast time $\tau_0$. However, compared to the zero order of complex amplitude $\varphi_{10}$, one can realize that the complex amplitude $\varphi_{20}$ may perform dynamical behavior on the fast time $\tau_0$. According to that, the primary system and the NES may occur oscillations on different time scales, which may lead to occurrence of the multiscale energy transfers.

Eq.(15) and Eq.(16) can be rewritten as the following equations:

$$\tau_0: \begin{cases} \dfrac{d\varphi_{10}}{d\tau_0} = 0 \\ \dfrac{d\varphi_{20}}{d\tau_0} = -(G_1 + G_2 + G_3) \end{cases} \quad (17)$$

and

$$\tau_1: \begin{cases} \dfrac{d\varphi_{10}}{d\tau_1} = -G_0 \\ \dfrac{d\varphi_{20}}{d\tau_1} = -\dfrac{1}{2}j\sigma\varphi_{20} \end{cases} \quad (18)$$

where

$$\begin{cases} G_0 = (\dfrac{\lambda_1}{2} + j\dfrac{\sigma}{\omega})\varphi_{10} + (\dfrac{\lambda_2}{2} - j\dfrac{k_2}{2\omega})(\varphi_{10} - \varphi_{20}) - \dfrac{A}{2} \\ G_1 = \dfrac{1}{2}j\varphi_{20} - (\dfrac{\lambda_2}{2} - j\dfrac{k_2}{2\omega})(\varphi_{10} - \varphi_{20}) + (\dfrac{\lambda}{8\omega^2} - j\dfrac{3k}{8\omega^3})\varphi_{20}^2\varphi_{20}^* \\ G_2 = G_1^* e^{-2j\omega\tau_0} \\ G_3 = -(\dfrac{\lambda}{8\omega^2} - j\dfrac{k}{8\omega^3})\varphi_{20}^3 e^{2j\omega\tau_0} - (\dfrac{\lambda}{8\omega^2} + j\dfrac{k}{8\omega^3})\varphi_{20}^{*3} e^{-4j\omega\tau_0} \end{cases} \quad (19)$$

The formula $G_0$ expresses the relationship between the amplitude $A$ of external excitation and the complex amplitudes $\varphi_{10}$ and $\varphi_{20}$ on the slow time $\tau_1$.

The equation $G_1=0$ is named the invariant manifold that approximately captures the dynamical behavior of the system (Eq.(5)) when the parameter $\varepsilon$ ($0<\varepsilon<<1$) is very small.

The formula $G_2$ is made up of $G_1^*$ produced by $e^{-2\omega\tau_0}$. And $G_1^*$ is the complex conjugate of $G_1$. This means the property of $G_2$ is equivalent to $G_1$. Thus, the equation $G_2=0$ is named the conjugate invariant manifold of $G_1=0$.

The formula $G_3$ denotes the higher-order frequency term. It depends on the fast time $\tau_0$ and contains the NES parameters ($\lambda$, $k$), and the external forcing parameter $\omega$ ($\omega=1+\varepsilon\sigma$), where $\omega$ represents the ratio of the external excitation frequency to the natural frequency of the primary system.

Viewed from Eq.(17) ~ (19), the complex amplitude $\varphi_{10}$ depends only on the slow time scale $\tau_1$: $\varphi_{10}=\varphi_{10}(\tau_1)$, and the complex amplitude $\varphi_{20}$ depends on the two time scales $\tau_0$ and $\tau_1$: $\varphi_{20}=\varphi_{20}(\tau_0,\tau_1)$. This means complex amplitude $\varphi_{10}$ contains the slow phase $\theta_{10}(\tau_1)$, while complex amplitude $\varphi_{20}$ may contain the two phases, namely the slow phase $\theta_S(\tau_1)$ and the fast phase $\theta_F(\tau_0)$. These phases imply the occurrence of frequencies on different time scales.

To facilitate the study of slow phases, $\theta_{10}(\tau_1)$ and $\theta_S(\tau_1)$, and fast phase, $\theta_F(\tau_0)$, we introduce the following assumptions:

***Assumption 1***: When the invariant manifold $G_1=0$ has a folding structure, all branches of the invariant manifold contain slow phases ($\theta_{10}(\tau_1)$, $\theta_S(\tau_1)$), and only the subset of the branches contain fast phase ($\theta_F(\tau_0)$).

***Assumption 2***: When the modulus of complex amplitude $\varphi_{20}$ is large enough to exceed a certain threshold value, the higher-order frequency term $G_3$ will be excited, which leads to the occurrence of fast phase $\theta_F(\tau_0)$ in $\varphi_{20}$.

### 3.2. Slow frequency of the complex amplitudes $\varphi_{10}$ and $\varphi_{20}$

According to *Assumption 1*, introducing the polar coordinates:

$$\begin{cases} \varphi_{10}(\tau_1) = N_{10}e^{j\theta_{10}(\tau_1)}, & N_{10} \in \mathbb{R}^+, \theta_{10} \in \mathbb{R} \\ \varphi_{20}(\tau_1) = N_{20}e^{j\theta_S(\tau_1)}, & N_{20} \in \mathbb{R}^+, \theta_S \in \mathbb{R} \end{cases} \quad (20)$$

and substituting Eq.(20) into $G_1 = 0$ we can acquire:

$$N_{10}e^{\theta_{10}(\tau_1)} = R_1 R_2 N_{20} e^{\theta_S(\tau_1)} \quad (21)$$

with

$$\begin{cases} R_1 = r_1 e^{\vartheta_1} \\ r_1 = \left(\lambda_2^2 + \dfrac{k_2^2}{\omega^2}\right)^{-\frac{1}{2}} \\ \vartheta_1 = \arctan\left(\dfrac{k_2}{\omega \lambda_2}\right) \end{cases} \quad (22)$$

and

$$\begin{cases} R_2 = r_2 e^{\vartheta_2} \\ r_2 = \left(\left(\lambda_2 + \dfrac{\lambda N_{20}^2}{4\omega^2}\right)^2 + \left(\dfrac{\omega - k_2}{\omega} - \dfrac{3kN_{20}^2}{4\omega^3}\right)^2\right)^{\frac{1}{2}} \\ \vartheta_2 = \arctan\left(\left(\dfrac{\omega - k_2}{\omega} - \dfrac{3kN_{20}^2}{4\omega^3}\right)\left(\lambda_2 + \dfrac{\lambda N_{20}^2}{4\omega^2}\right)^{-1}\right) \end{cases} \quad (23)$$

Therefore, we get the phase difference between the complex amplitudes $\varphi_{10}$ and $\varphi_{20}$:

$$\Delta = \theta_{10}(\tau_1) - \theta_S(\tau_1) = \vartheta_1 + \vartheta_2 \quad (24)$$

Furthermore, in Eq.(18), we can get:

$$\varphi_{20}(\tau_1) = Ce^{-j\frac{1}{2}\sigma\tau_1} \quad (25)$$

The frequency $\omega_S$ of complex amplitude $\varphi_{20}$ on the slow time scale $\tau_1$ is solved:

$$\omega_{\mathrm{S}} = \frac{1}{2}\sigma \tag{26}$$

Eq.(24) indicates the two slow phases, $\theta_{10}(\tau_1)$, and $\theta_{\mathrm{S}}(\tau_1)$, contain equal slow frequency, namely $\omega_{\mathrm{S}}$, since $\vartheta_1$ and $\vartheta_2$ both are constant. Thus, we obtain:

$$\frac{d\theta_{10}(\tau_1)}{d\tau_1} = \frac{d\theta_{\mathrm{S}}(\tau_1)}{d\tau_1} = \omega_{\mathrm{S}} = \frac{1}{2}\sigma \tag{27}$$

### 3.3. Fast frequency of the complex amplitude $\varphi_{20}$

The higher-order frequency term $G_3$ is rewritten as:

$$G_3 = -(V + V^*)e^{-\mathrm{j}\omega\tau_0} \tag{28}$$

with

$$\begin{cases} V = (\dfrac{\lambda}{8\omega^2} - \mathrm{j}\dfrac{k}{8\omega^3})\varphi_{20}^3 e^{3\mathrm{j}\omega\tau_0} \\ V^* = (\dfrac{\lambda}{8\omega^2} + \mathrm{j}\dfrac{k}{8\omega^3})\varphi_{20}^{*3} e^{-3\mathrm{j}\omega\tau_0} \end{cases} \tag{29}$$

where $V$ and $V^*$ are the complex conjugate to each other.

For $G_3=0$, one can get:

$$V + V^* = 0 \tag{30}$$

When the higher-order frequency term $G_3$ is excited, we introduce the following polar coordinate:

$$\varphi_{20}(\tau_0, \tau_1) = N_{20} e^{\mathrm{j}\theta_{\mathrm{S}}(\tau_1)} e^{\mathrm{j}\theta_{\mathrm{F}}(\tau_0)}, \, N_{20} \in \mathbb{R}^+, \theta_{\mathrm{S}} \in \mathbb{R}, \theta_{\mathrm{F}} \in \mathbb{R} \tag{31}$$

The part $N_{20}e^{\mathrm{j}\theta_{\mathrm{S}}}$, produced by the amplitude $N_{20}$ and the slow phase $\theta_{\mathrm{S}}$, needs to satisfy $G_1 = 0$. Therefore, we get the following equations:

$$\begin{cases} V = Re^{\mathrm{j}\vartheta} \\ V^* = R^* e^{-\mathrm{j}\vartheta} \end{cases} \tag{32}$$

where

$$\begin{cases} R = \left(\dfrac{\lambda^2}{64\omega^4} + \dfrac{k^2}{64\omega^6}\right)^{\frac{1}{2}} \\ \vartheta = 3\omega\tau_0 + 3\theta_{\mathrm{F}}(\tau_0) + \delta \\ \delta = \arctan\left(-\dfrac{k}{\omega\lambda}\right) \end{cases} \tag{33}$$

When $V+V^*=0$, one can obtain:

$$\mathrm{Re}(V) = R\cos\vartheta = 0 \tag{34}$$

with

$$\vartheta = 3\omega\tau_0 + 3\theta_F(\tau_0) + \delta = \pm\frac{\pi}{2} \pm n\pi, \quad (n = 0,1,2,3...) \tag{35}$$

According to that, the fast phase $\theta_F(\tau_0)$ of the complex amplitude $\varphi_{20}$ is solved:

$$\theta_F(\tau_0) = -\omega\tau_0 + \frac{1}{3}\left(\pm\frac{\pi}{2} \pm n\pi - \delta\right), \quad (n = 0,1,2,3...) \tag{36}$$

Consequently, the fast frequency $\omega_F$ corresponding to the fast phase $\theta_F(\tau_0)$ is got:

$$\omega_F = \left|\frac{d\theta_F(\tau_0)}{d\tau_0}\right| = \omega \tag{37}$$

### 3.4. Invariant manifold

Letting $G_1=0$, we get the invariant manifold $\Pi$:

$$\Pi = \left\{(\varphi_{10}, \varphi_{20}) \in \mathbb{C}^2 : G_1 = \frac{1}{2}j\varphi_{20} - \left(\frac{\lambda_2}{2} - j\frac{k_2}{2\omega}\right)(\varphi_{10} - \varphi_{20}) + \left(\frac{\lambda}{8\omega^2} - j\frac{3k}{8\omega^3}\right)|\varphi_{20}|^2 \varphi_{20} = 0\right\} \tag{38}$$

Substituting Eq.(20) into Eq.(38) and putting $Z_{10}=N_{10}^2$ and $Z_2=N_{20}^2$, one can obtain the cubic polynomial:

$$a_3 Z_{20}^3 + a_2 Z_{20}^2 + a_1 Z_{20} + a_0 Z_{10} = 0 \tag{39}$$

where

$$\begin{cases} a_0 = -\left(\lambda_2^2 \omega^2 + k_2^2\right) \\ a_1 = \left(\lambda_2^2 + 1\right)\omega^2 + \left(k_2 - 2\omega\right)k_2 \\ a_2 = \frac{1}{2\omega^2}\left(\lambda_2 \lambda \omega^2 + 3kk_2 - 3k\omega\right) \\ a_3 = \frac{1}{16\omega^4}\left(\lambda^2 \omega^2 + 9k^2\right) \end{cases} \tag{40}$$

Viewed in Eq.(39), the number of roots for $Z_{20}$ will change by changing $Z_{10}$, implying the occurrence of a folding kind of bifurcation point, namely saddle-node (SN) bifurcation. As the value of $Z_{10}$ and other system parameters are fixed at adequate values, Eq.(39) will have one or three positive roots with respect to $Z_{20}$. Letting the derivative of Eq. (39) with respect to $Z_{20}$ equals zero, the bifurcation condition is obtained:

$$3a_3 Z_{20}^2 + 2a_2 Z_{20} + a_1 = 0 \tag{41}$$

When coefficients $a_i$ ($i=0,1,2,3$) satisfy the condition $a_2^2 - 3a_1 a_3 > 0$, there will be a pair of SN bifurcation points with the change of $Z_{10}$. The two SN bifurcation points are noted as $SN_1$ and $SN_2$, respectively.

For simplicity, we project the invariant manifold onto its modulus plane ($N_{10}$, $N_{20}$). Fixing the system parameters at $\varepsilon=0.001$, $k=0.5$, $k_2=0.5$, $\lambda=0.1$, $\lambda_1=0.1$, $\lambda_2=0.1$, and $\sigma=3.14$, we can get the invariant manifold on modulus plane ($N_{10}$, $N_{20}$), depicted in Fig.2.. There is a folding structure on the invariant manifold. The folding structure consists of two stable invariant manifolds, namely the lower branch and the upper branch, and one unstable invariant manifold, namely the middle branch. According to *Assumption 1*, all branches of the invariant manifold contain the slow frequency $\omega_S$. In Fig.2., one can notice that the modulus $N_{20}$, meeting the bifurcation points $SN_1$ and $SN_2$, will occur jumps with the change of the modulus $N_{10}$. Moreover, it is necessary to consider that only when the energy of the NES is large enough will its complex amplitude oscillate rapidly. Therefore, combined with *Assumption 2*, we assume that there is a fast frequency $\omega_F$ in the upper branch of the invariant manifold.

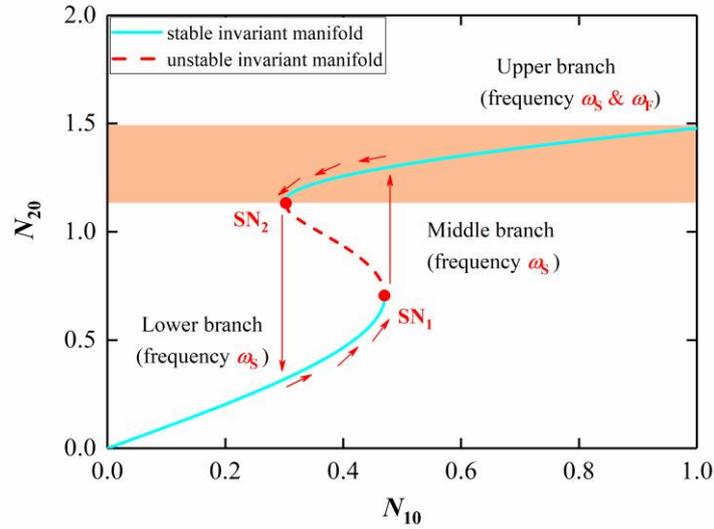

Fig.2. The invariant manifold on the modulus plane ($N_{10}$, $N_{20}$) by fixing system parameters at $\varepsilon=0.001$, $k=0.5$, $k_2=0.5$, $\lambda=0.1$, $\lambda_1=0.1$, $\lambda_2=0.1$, and $\sigma=3.14$.

### 3.5. Amplitude-frequency diagram

Amplitude-frequency curves with the change of the amplitude $A$ of external forcing are obtained by fixing system parameters at $\varepsilon=0.001$, $k=0.5$, $k_2=0.5$, $\lambda=0.1$, $\lambda_1=0.1$, $\lambda_2=0.1$. In Fig.3., the maximum of $N_{20}$ gradually increases by increasing the external forcing amplitude $A$.

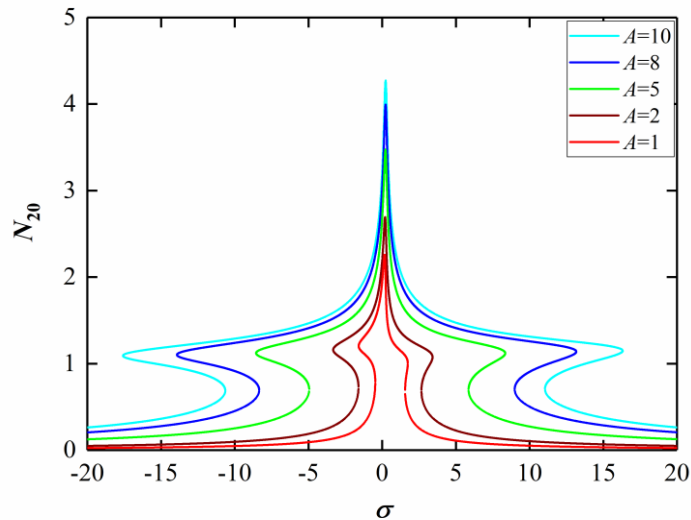

Fig.3. Amplitude-frequency curves for the NES with the change of the amplitude $A$ of external forcing by fixing system parameters at $\varepsilon=0.001$, $k=0.5$, $k_2=0.5$, $\lambda=0.1$, $\lambda_1=0.1$, $\lambda_2=0.1$.

Moreover, when a pair of SN bifurcation points occur, the thresholds of the amplitude $A$ of external forcing can be obtained from the following equation as the parameter $\sigma$ and the other system parameters are fixed.

$$b_3 Z_{20}^3 + b_2 Z_{20}^2 + b_1 Z_{20} + b_0 A^2 = 0 \tag{42}$$

where

$$\begin{cases} b_0 = -16\omega^6 \left(\lambda_2^2 \omega^2 + k_2^2\right) \\ b_1 = 16\omega^4 \begin{pmatrix} 4k_2^2 \sigma^2 + 4k_2 \sigma(-2\sigma + k_2)\omega + \left(4\left(\lambda_2^2 + 1\right)\sigma^2 - 4k_2 \sigma + k_2^2\left(\lambda_1^2 + 1\right)\right)\omega^2 \\ +2\left(-k_2 \lambda_1^2 + 2\sigma \lambda_2^2\right)\omega^3 + \left(\left(\lambda_2^2 + 1\right)\lambda_1^2 + 2\lambda_1 \lambda_2 + \lambda_2^2\right)\omega^4 \end{pmatrix} \\ b_2 = -24\omega^2 \begin{pmatrix} 2kk_2 \sigma(-2\sigma + k_2) + k(-2\sigma + k_2)^2 \omega + \left(\left(-k_2 \lambda_1^2 + 2\sigma \lambda_2^2\right)k - \frac{1}{3}\lambda\left(k_2^2 \lambda_1 + 4\lambda_2 \sigma^2\right)\right)\omega^2 \\ +k(\lambda_1 + \lambda_2)^2 \omega^3 - \frac{1}{3}\lambda \lambda_1 \lambda_2 (\lambda_1 + \lambda_2)\omega^4 \end{pmatrix} \\ b_3 = \left(\omega^2 \lambda^2 + 9k^2\right)\left((\lambda_1 + \lambda_2)^2 \omega^2 + (-2\sigma + k_2)^2\right) \end{cases} \tag{43}$$

The bifurcation condition is obtained:

$$3b_3 Z_{20}^2 + b_2 Z_{20} + b_1 = 0 \tag{44}$$

In Fig.4., the two thresholds $A_{C1}$ and $A_{C2}$ divide the bifurcation diagram into three regions. When A< $A_{C2}$ and A> $A_{C1}$, there is only one fixed point. However, when $A_{C2}$< A < $A_{C1}$, there are three fixed points. As the amplitude $A$ meets the thresholds, the stability of $N_{20}$ will be changed.

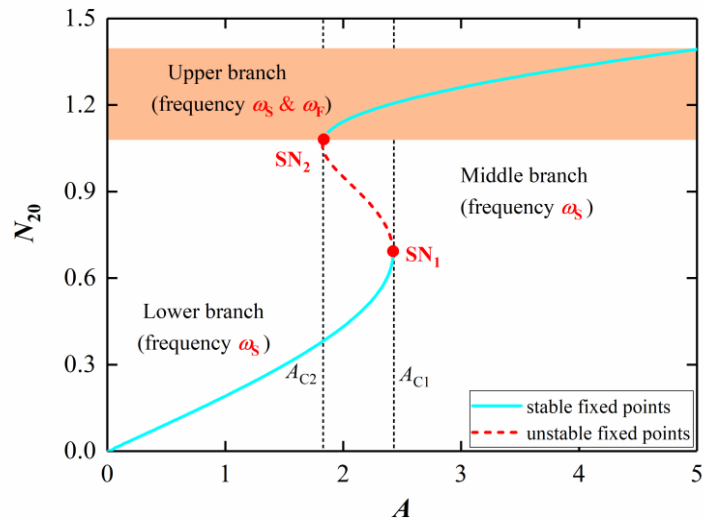

Fig.4. Bifurcation diagram by fixing system parameters at $\varepsilon$=0.001, $k$=0.5, $k_2$=0.5, $\lambda$=0.1, $\lambda_1$=0.1, $\lambda_2$=0.1, and $\sigma$=3.14.

# 4. Classification of the steady-state oscillations in the system

## 4.1. Determination of the steady-state oscillations

To discuss and classify the steady-state oscillations of system, Eq.(15) and Eq.(16) are rewritten as:

$$\tau_0: \begin{cases} \dfrac{d\varphi_{10}}{d\tau_0} = 0 \\ \dfrac{d\varphi_{20}}{d\tau_0} = -(G_1' + G_2' + G_3) \end{cases} \quad (45)$$

and

$$\tau_1: \begin{cases} \dfrac{d\varphi_{10}}{d\tau_1} = -G_0 \\ \dfrac{d\varphi_{20}}{d\tau_1} = 0 \end{cases} \quad (46)$$

with

$$\begin{cases} G_0 = (\dfrac{\lambda_1}{2} + j\dfrac{\sigma}{\omega})\varphi_{10} + (\dfrac{\lambda_2}{2} - j\dfrac{k_2}{2\omega})(\varphi_{10} - \varphi_{20}) - \dfrac{A}{2} \\ G_1' = \dfrac{1}{2}j\omega\varphi_{20} - (\dfrac{\lambda_2}{2} - j\dfrac{k_2}{2\omega})(\varphi_{10} - \varphi_{20}) + (\dfrac{\lambda}{8\omega^2} - j\dfrac{3k}{8\omega^3})\varphi_{20}^2\varphi_{20}^* \\ G_2' = G_1'^* e^{-2j\omega\tau_0} \\ G_3 = -(\dfrac{\lambda}{8\omega^2} - j\dfrac{k}{8\omega^3})\varphi_{20}^3 e^{2j\omega\tau_0} - (\dfrac{\lambda}{8\omega^2} + j\dfrac{k}{8\omega^3})\varphi_{20}^{*3} e^{-4j\omega\tau_0} \end{cases} \quad (47)$$

When the system (Eq.(45)~Eq.(47)) satisfies the conditions ($G_0 = 0, G_1' + G_2' + G_3 = 0$) several types of steady-state oscillations of the system will occur. Note that the two system (Eq.(45)~Eq.(47)) and (Eq.(17)~Eq.(19)) are equivalent. Moreover, by fixing the system parameters and the external forcing parameters at different values, the system may have distinct types of steady-state oscillations, e.g., stationary oscillations, nonlinear beats, and relaxation oscillations. These oscillations exhibit very different properties.

For the stationary oscillations, the system response amplitudes do not vary with time. That is, the complex amplitudes $\varphi_1$ and $\varphi_2$ do not exhibit any time scale variation and do not trigger the higher-order frequency term $G_3$. For the nonlinear beats, the complex amplitudes $\varphi_1$ and $\varphi_2$ exhibit variation on the slow time scale $\tau_1$, meaning that the system response amplitudes vary slowly. Similarly, the nonlinear beats also do not trigger the higher-order frequency term $G_3$. However, for the relaxation oscillations, the amplitudes of the system responses not only vary on the slow time $\tau_1$ but also trigger the higher-order frequency term $G_3$, resulting in rapid oscillations on the fast time $\tau_0$.

It should be noted that the three types of oscillations, namely the stationary oscillations, the nonlinear beats, and the relaxation oscillations, are all steady-state oscillations. Therefore, the complexification systems

of these oscillations all satisfy the following equations:

$$\tau_0: \begin{cases} \dfrac{d\varphi_{10}}{d\tau_0} = 0 \\ \dfrac{d\varphi_{20}}{d\tau_0} = 0 \end{cases}$$
$$\tau_1: \begin{cases} \dfrac{d\varphi_{10}}{d\tau_1} = 0 \\ \dfrac{d\varphi_{20}}{d\tau_1} = 0 \end{cases} \quad (48)$$

**4.2. Stationary oscillations**

When the stationary oscillations occur, the system can be expressed in the following form:

$$\tau_0: \begin{cases} \dfrac{d\varphi_{10}}{d\tau_0} = 0 \\ \dfrac{d\varphi_{20}}{d\tau_0} = -(G'_1 + G'_2) \end{cases} \quad (49)$$

and

$$\tau_1: \begin{cases} \dfrac{d\varphi_{10}}{d\tau_1} = -G_0 \\ \dfrac{d\varphi_{20}}{d\tau_1} = 0 \end{cases} \quad (50)$$

with

$$\begin{cases} G_0 = (\dfrac{\lambda_1}{2} + j\dfrac{\sigma}{\omega})\varphi_{10} + (\dfrac{\lambda_2}{2} - j\dfrac{k_2}{2\omega})(\varphi_{10} - \varphi_{20}) - \dfrac{A}{2} \\ G'_1 = \dfrac{1}{2}j\omega\varphi_{20} - (\dfrac{\lambda_2}{2} - j\dfrac{k_2}{2\omega})(\varphi_{10} - \varphi_{20}) + (\dfrac{\lambda}{8\omega^2} - j\dfrac{3k}{8\omega^3})\varphi_{20}^2\varphi_{20}^* \\ G'_2 = G'^*_1 e^{-2j\omega\tau_0} \end{cases} \quad (51)$$

Fixing the system parameters and the external forcing parameters at $\varepsilon=0.001$, $k=0.5$, $k_2=0.5$, $\lambda=0.1$, $\lambda_1=0.1$, $\lambda_2=0.1$, $\sigma=6$ and $A=1$, there are the stationary oscillations in the system because of satisfying Eq.(48). One can notice that the numerical result of amplitudes of system responses is a fixed point, denoted as $p_0$. This fixed point $p_0$ is captured by the invariant manifold projected on the modulus plane ($N_{10}$, $N_{20}$), depicted in subfigure (a) of Fig.5. The fixed point stands for the stationary oscillation, and represents a cycle on the phase portrait of the system, seeing in subfigure (b) of Fig.5. Thus, the time series of amplitudes of system responses keep unchanged except for a short time in the beginning, viewed in subfigures (c) and (b) of Fig.5. Note that the stationary oscillations do not trigger the slow frequency $\omega_S$, just as it is captured by the invariant manifold as a fixed point.

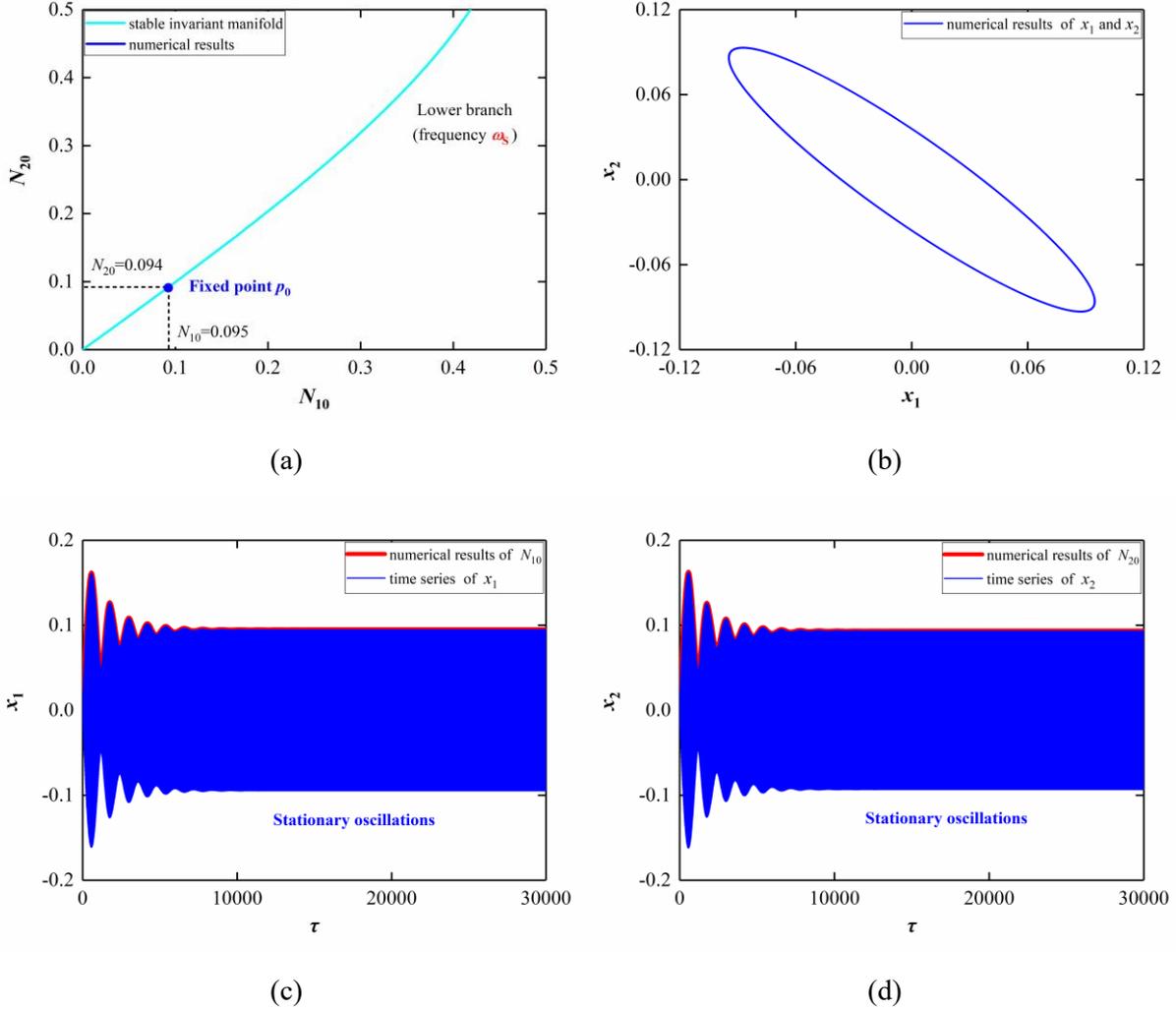

Fig.5. Numerical simulations with the system parameters fixing at $\varepsilon=0.001$, $k=0.5$, $k_2=0.5$, $\lambda=0.1$, $\lambda_1=0.1$, $\lambda_2=0.1$, $\sigma=6$ and $A=1$. (a) Superposition of the invariant manifold on the modulus plane ($N_{10}$, $N_{20}$) and corresponding numerical results;(b) Phase portrait; (c) Superposition of the time series for the displacement $x_1$ of the primary system and the complex amplitude $\varphi_1$; (d) Superposition of the time series for the displacement $x_2$ of NES and the complex amplitude $\varphi_2$.

### 4.3. Nonlinear beats

When the nonlinear beats occur, the complexification system can be rewritten as the following equations:

$$\tau_0: \begin{cases} \dfrac{d\varphi_{10}}{d\tau_0} = 0 \\ \dfrac{d\varphi_{20}}{d\tau_0} = -(G_1 + G_2) \end{cases} \qquad (52)$$

and

$$\tau_1: \begin{cases} \dfrac{d\varphi_{10}}{d\tau_1} = -G_0 \\ \dfrac{d\varphi_{20}}{d\tau_1} = -\dfrac{1}{2}j\sigma\varphi_{20} \end{cases} \qquad (53)$$

with

$$\begin{cases} G_0 = (\dfrac{\lambda_1}{2}+j\dfrac{\sigma}{\omega})\varphi_{10} + (\dfrac{\lambda_2}{2}-j\dfrac{k_2}{2\omega})(\varphi_{10}-\varphi_{20}) - \dfrac{A}{2} \\ G_1 = \dfrac{1}{2}j\varphi_{20} - (\dfrac{\lambda_2}{2}-j\dfrac{k_2}{2\omega})(\varphi_{10}-\varphi_{20}) + (\dfrac{\lambda}{8\omega^2}-j\dfrac{3k}{8\omega^3})\varphi_{20}^2\varphi_{20}^* \\ G_2 = G_1^* e^{-2j\omega\tau_0} \end{cases} \quad (54)$$

Fix the system parameters and the external forcing parameters at $\varepsilon=0.001$, $k=0.5$, $k_2=0.5$, $\lambda=0.1$, $\lambda_1=0.1$, $\lambda_2=0.1$, $\sigma=3.5$ and $A=1$, and the system will exhibit nonlinear beats since it still satisfies Eq.(48). The numerical results of the amplitudes of system responses exhibit an interval, denoted as the interval $[p_1, p_2]$, on modulus plane $(N_{10}, N_{20})$. It is captured by the invariant manifold, seen in subfigure (a) of Fig.6.. The trajectory of the nonlinear beat shows a ring on the phase portrait of the system, depicted in subfigure (b) of Fig.6.. It is necessary to further clarify that interval $[p_1, p_2]$ represents the periodic change on the slow time $\tau_1$ in the amplitudes of the system responses. In subfigures (c) and (d) of Fig.6., we can find the time series of the system responses containing the slowly-varying amplitudes. Hence, the numerical simulations verify that the slow frequency $\omega_S$ is triggered by the nonlinear beats.

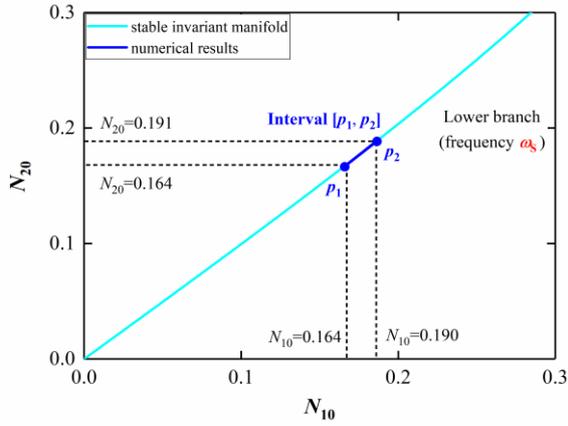

(a)

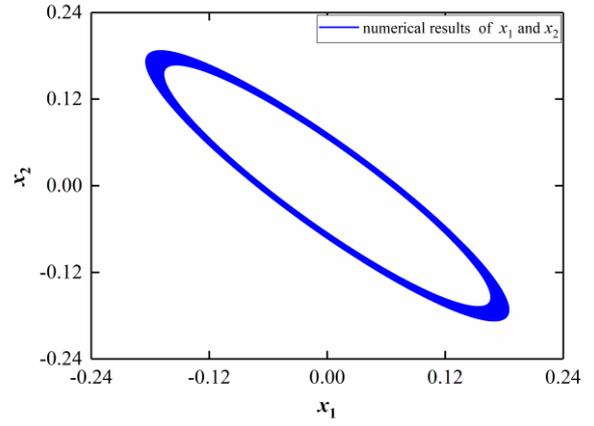

(b)

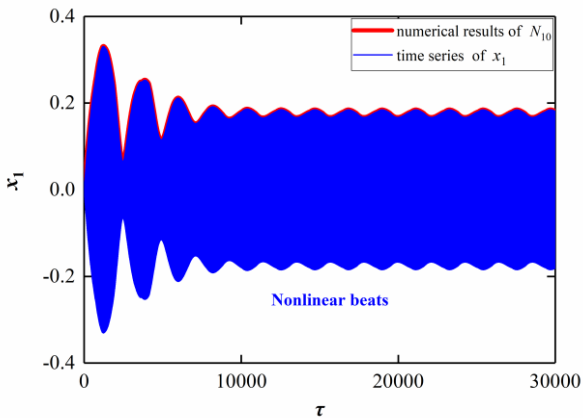

(c)

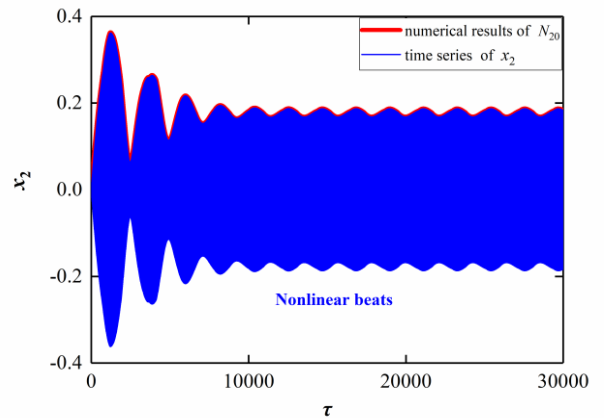

(d)

Fig.6. Numerical simulations with the system parameters fixing at $\varepsilon=0.001$, $k=0.5$, $k_2=0.5$, $\lambda=0.1$, $\lambda_1=0.1$, $\lambda_2=0.1$, $\sigma=3.5$ and $A=1$. (a) Superposition of the invariant manifold on the modulus plane ($N_{10}$, $N_{20}$) and corresponding numerical results;(b) Phase portrait; (c) Superposition of the time series for the displacement $x_1$ of the primary system and the complex amplitude $\varphi_1$; (d) Superposition of the time series for the displacement $x_2$ of NES and the complex amplitude $\varphi_2$.

According to Eq.(27), the slow frequency is predicted:

$$\omega_S = \frac{1}{2}\sigma = 1.75 \tag{55}$$

Then, we get the theoretical predicted period $T_S$:

$$T_S = \frac{\pi}{\omega_S} \times \frac{1}{\varepsilon} \approx 1800 \tag{56}$$

Viewed in subfigures (a) and (b) of Fig.7., the numerical results show the periods of the slowly-varying amplitudes for system responses:

$$T_1 = T_2 \approx 2150 \tag{57}$$

Furthermore, the corresponding frequencies, $\omega_1$, $\omega_2$, are numerically got:

$$\omega_1 = \omega_2 = \frac{\pi}{\varepsilon T_1} = \frac{\pi}{\varepsilon T_2} \approx 1.46 \tag{58}$$

By comparing Eq.(55) and Eq.(58), or comparing Eq.(56) and Eq.(57), we can find that the system (Eq.(52)~(54)) can well describe the nonlinear beats phenomenon, since the slow frequency ($\omega_S=\sigma/2$) solved by Eq.(52)~(54) predicts the frequencies, $\omega_1$, $\omega_2$ of the slowly-varying amplitudes of system responses approximately.

There may be a weak energy transfer occurring in the nonlinear beats since the response amplitudes of the primary system and the NES both are slowly-varying.

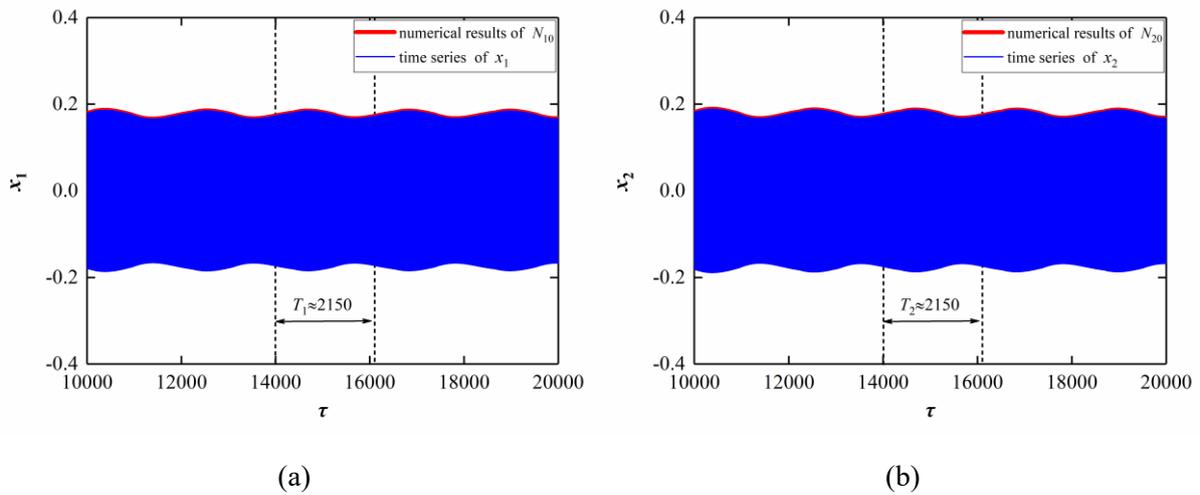

(a)        (b)

Fig.7. Numerical simulations with the system parameters fixing at $\varepsilon=0.001$, $k=0.5$, $k_2=0.5$, $\lambda=0.1$, $\lambda_1=0.1$, $\lambda_2=0.1$, $\sigma=3.5$ and $A=1$. (a) Close up 1;(b) Close up 2.

### 4.4. Relaxation oscillations

When the relaxation oscillations occur, the complexification system can be rewritten as:

$$\tau_0: \begin{cases} \dfrac{d\varphi_{10}}{d\tau_0} = 0 \\ \dfrac{d\varphi_{20}}{d\tau_0} = -(G_1 + G_2 + G_3) \end{cases} \quad (59)$$

and

$$\tau_1: \begin{cases} \dfrac{d\varphi_{10}}{d\tau_1} = -G_0 \\ \dfrac{d\varphi_{20}}{d\tau_1} = -\dfrac{1}{2} j\sigma\varphi_{20} \end{cases} \quad (60)$$

with

$$\begin{cases} G_0 = (\dfrac{\lambda_1}{2} + j\dfrac{\sigma}{\omega})\varphi_{10} + (\dfrac{\lambda_2}{2} - j\dfrac{k_2}{2\omega})(\varphi_{10} - \varphi_{20}) - \dfrac{A}{2} \\ G_1 = \dfrac{1}{2} j\varphi_{20} - (\dfrac{\lambda_2}{2} - j\dfrac{k_2}{2\omega})(\varphi_{10} - \varphi_{20}) + (\dfrac{\lambda}{8\omega^2} - j\dfrac{3k}{8\omega^3})\varphi_{20}^2 \varphi_{20}^* \\ G_2 = G_1^* e^{-2j\omega\tau_0} \\ G_3 = -(\dfrac{\lambda}{8\omega^2} - j\dfrac{k}{8\omega^3})\varphi_{20}^3 e^{2j\omega\tau_0} - (\dfrac{\lambda}{8\omega^2} + j\dfrac{k}{8\omega^3})\varphi_{20}^{*3} e^{-4j\omega\tau_0} \end{cases} \quad (61)$$

#### 4.4.1. Case 1: σ=3.14

Taking the system parameters and the external forcing parameters at $\varepsilon=0.001$, $k=0.5$, $k_2=0.5$, $\lambda=0.1$, $\lambda_1=0.1$, $\lambda_2=0.1$, $\sigma=3.14$ and $A=10$, we can notice that there are the hysteresis loops on modulus plane ($N_{10}$, $N_{20}$). In this case, the system triggers the high-order frequency term $G_3$ and satisfies Eq.(48). The numerical results show that the amplitude of NES response makes a jump when it meets the bifurcation points $SN_1$ or $SN_2$, and performs the fast oscillations on the upper branch of the invariant manifold. The detailed information can be seen in subfigures (a) and (d) of Fig.8. Note that the primary system responses show nonlinear beats.

Similar to Section 4.3, we get the theoretical slow frequency $\omega_S$:

$$\omega_S = \dfrac{1}{2}\sigma = 1.57 \quad (62)$$

and the theoretical period $T_S$:

$$T_S = \dfrac{\pi}{\omega_S} \times \dfrac{1}{\varepsilon} \approx 2000 \quad (63)$$

In subfigures (a) and (b) of Fig.9., the periods of the slowly-varying amplitudes for system responses are obtained by numerical simulations. They are shown below:

$$T_1 = T_2 \approx 2100 \tag{64}$$

Then, the frequencies, $\omega_1$, $\omega_2$, are numerically got:

$$\omega_1 = \omega_2 = \frac{\pi}{\varepsilon T_1} = \frac{\pi}{\varepsilon T_2} \approx 1.50 \tag{65}$$

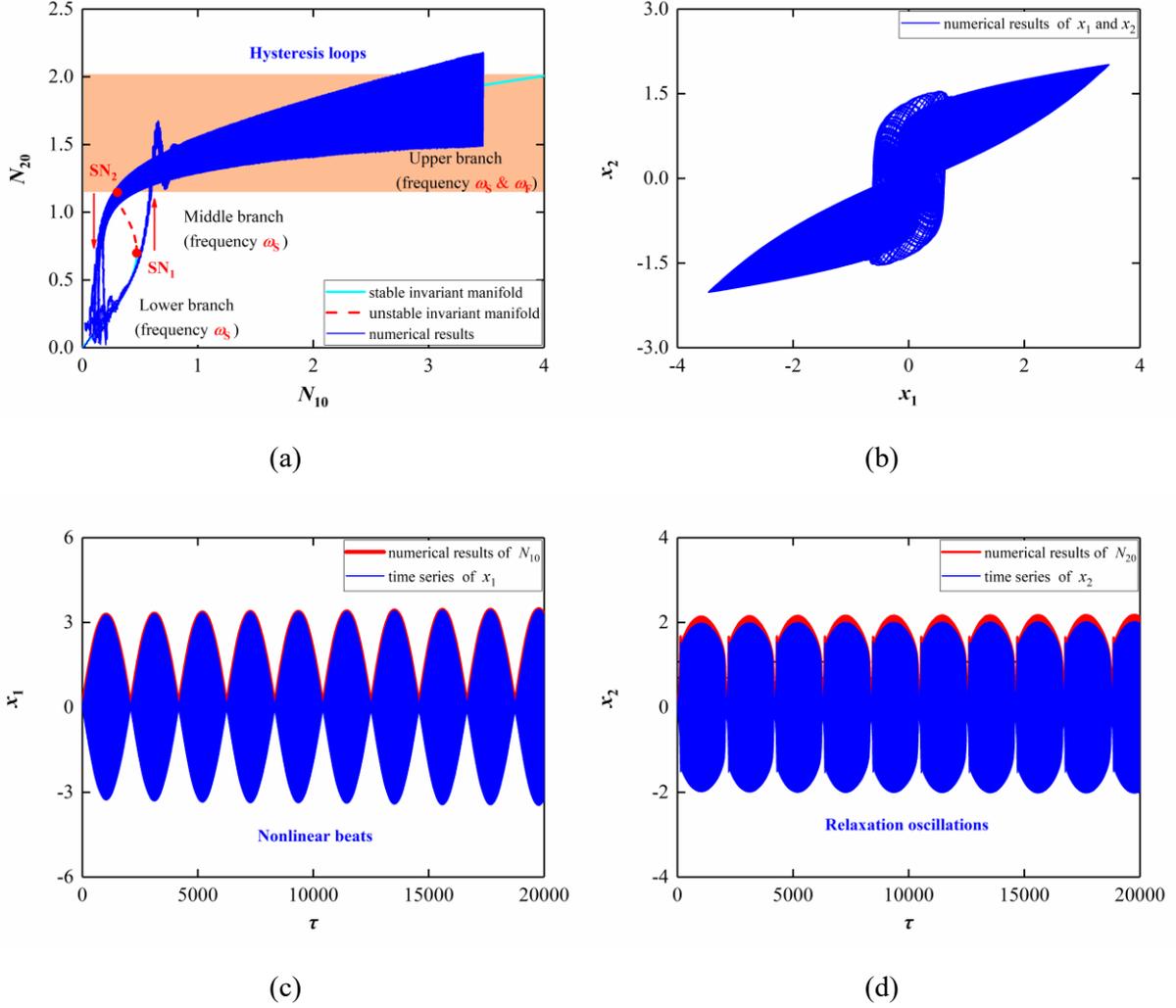

(a)            (b)

(c)            (d)

Fig.8. Numerical simulations with the system parameters fixing at $\varepsilon=0.001$, $k=0.5$, $k_2=0.5$, $\lambda=0.1$, $\lambda_1=0.1$, $\lambda_2=0.1$, $\sigma=3.14$ and $A=10$. (a) Superposition of the invariant manifold on the modulus plane $(N_{10}, N_{20})$ and corresponding numerical results; (b) Phase portrait; (c) Superposition of the time series for the displacement $x_1$ of the primary system and the complex amplitude $\varphi_1$; (d) Superposition of the time series for the displacement $x_2$ of NES and the complex amplitude $\varphi_2$.

Furthermore, since the relaxation oscillations trigger the high-order frequency term $G_3$, the fast frequency $\omega_F$ is solved by Eq.(37). Viewed from subfigure (c) of Fig.9., the amplitude of NES response makes twice jumping when it meets $N_{20}=0.693$ and $N_{20}=1.072$, respectively. Thus, the fast frequency $\omega_F$ is predicted:

$$\omega_F \approx 1 \tag{66}$$

The corresponding period $T_F$ is shown as:

$$T_F = \frac{\pi}{\omega_F} \approx 3 \tag{67}$$

In subfigure (d) of Fig.9., the numerical results show one period of the fast oscillation, captured by the upper branch of the invariant manifold, is equal to $T_3 \approx 3$. Thus, the fast frequency $\omega_3$ is numerically obtained:

$$\omega_3 = \frac{\pi}{T_3} \approx 1 \tag{68}$$

In this case, it should be noted that the response of the primary system is represented by the nonlinear beat, while the NES response is represented by the relaxation oscillation, implying the occurrence of energy transfers on different time scales. Moreover, there is the obvious fact that the slow frequency $\omega_S$, obtained by theoretical analysis, is close to the frequencies, $\omega_1$, and $\omega_2$, obtained by the numerical results. Additionally, the frequencies, $\omega_F$, and $\omega_3$ are almost equal to each other. This indicates that the complexification system (Eq.(59)~(61)) is valid.

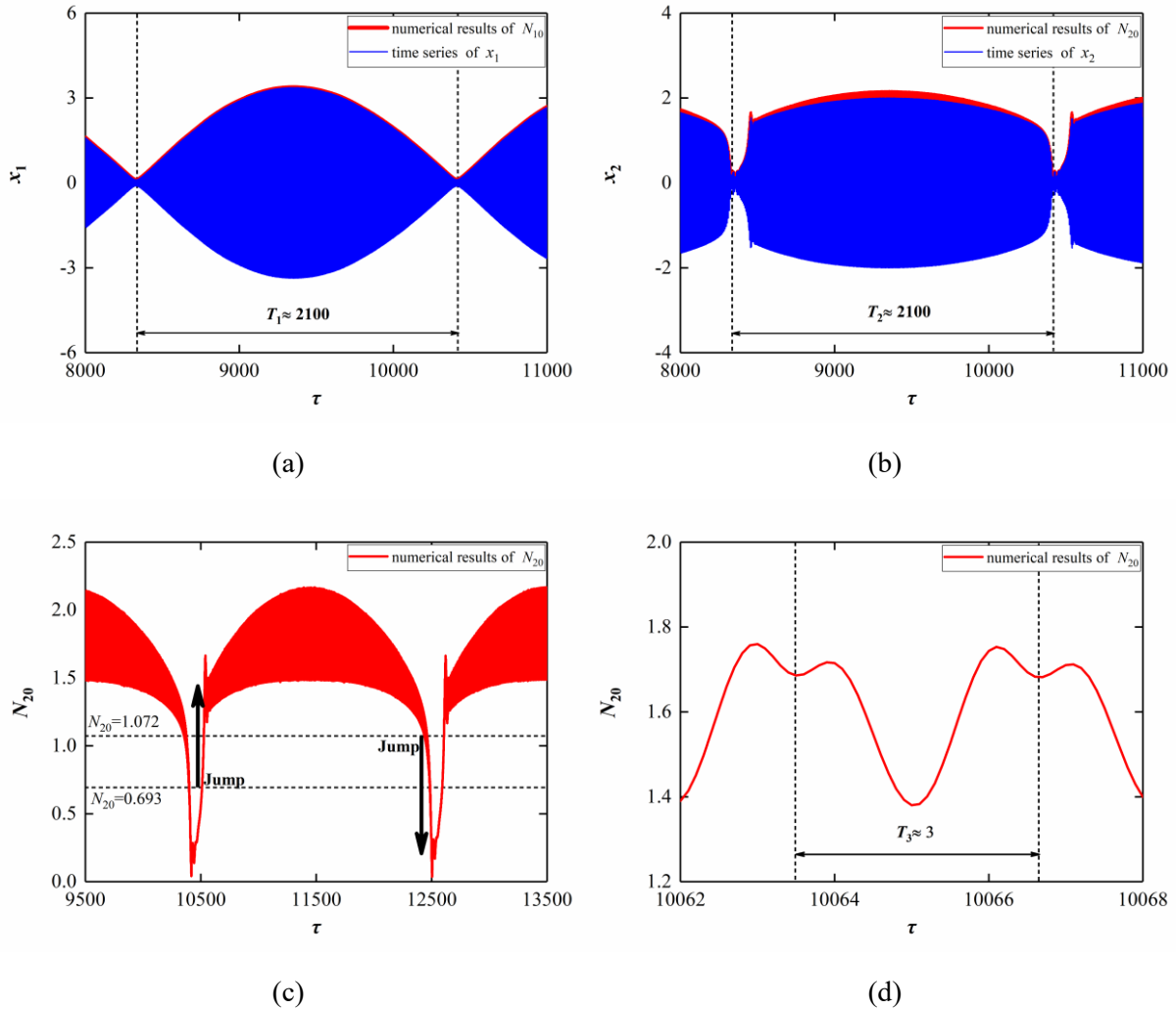

Fig.9. Numerical simulations with the system parameters fixing at $\varepsilon=0.001$, $k=0.5$, $k_2=0.5$, $\lambda=0.1$, $\lambda_1=0.1$, $\lambda_2=0.1$, $\sigma=3.14$ and $A=10$. (a) Close up 1;(b) Close up 2; (c) Close up 3;(d) Close up 4.

### 4.4.2. Case 2: $\sigma=2$

In Case 1, the parameter $\sigma$ is fixed at 3.14 which is close to $\pi$. This seems a bit special. Without losing generality, we here fix $\sigma$ at other values as a reference.

Take the value of the parameter $\sigma$ at 2 and keep the other system parameters unchanged.

$$\omega_s = \frac{1}{2}\sigma = 1 \tag{69}$$

The system performs the relaxation oscillations, seeing subfigure (b) of Fig.(10). Similar to Case 1, Case 2 triggers the high-order frequency term $G_3$ and satisfies Eq.(48). For the sake of simplicity, we directly give the slow frequency, $\omega_1$, $\omega_2$, obtained from the numerical results. They are:

$$\omega_1 = \omega_2 \approx 0.92 \tag{70}$$

In addition, one period of the fast oscillation, obtained by the numerical results, remains unchanged, i.e., $T_3 \approx 3$, depicted in subfigure (d) of Fig.(10).

This situation further validates the effectiveness of the complexification system (Eq.(59)~(61)) for describing the relaxation oscillations.

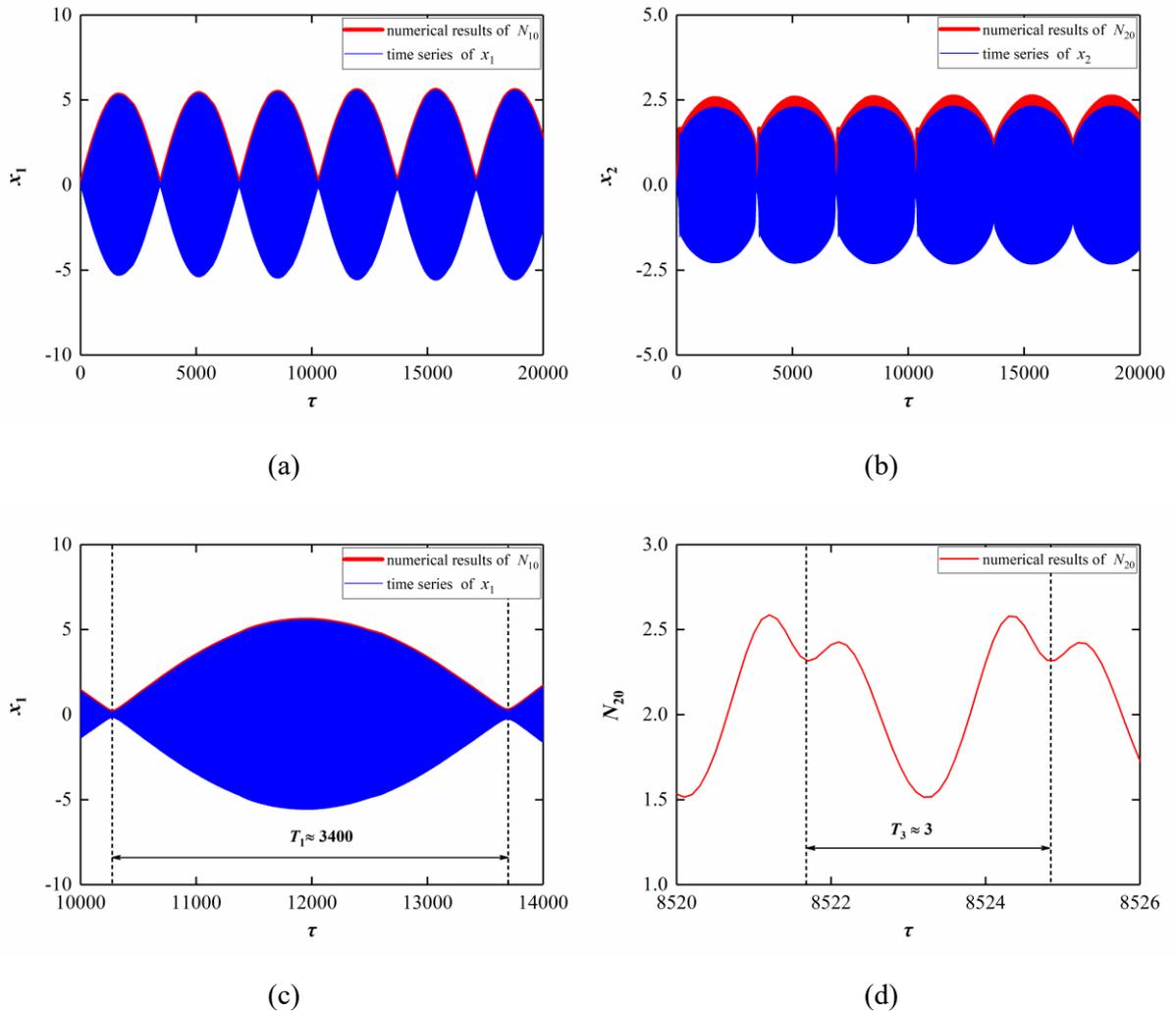

Fig.10. Numerical simulations with the system parameters fixing at $\varepsilon=0.001$, $k=0.5$, $k_2=0.5$, $\lambda=0.1$, $\lambda_1=0.1$, $\lambda_2=0.1$, $\sigma=2$ and $A=10$. (a) Superposition of the time series for the displacement $x_1$ of the primary system and the complex amplitude $\varphi_1$; (b) Superposition of the time series for the displacement $x_2$ of NES and the complex amplitude $\varphi_2$; (c) Close up 1; (d) Close up 2.

# 5. Energy transfers on different time scales

## 5.1. Determination of energy of the system

The energies of the primary system and the NES are represented as follows:

$$\begin{cases} E_{LO} = \frac{1}{2}\left(\dot{x}_1^2 + x_1^2\right) \\ E_{NES} = \frac{1}{2}\varepsilon\left(\dot{x}_2^2 + k_2(x_2 - x_1)^2 + \frac{1}{2}kx_2^4\right) \end{cases} \quad (71)$$

$E_{LO}$ stands for the energy of the primary system for the non-dimensional system (Eq.(5)). The mass and the stiffness of the primary system both are equal to 1. For the NES, its energy is expressed by $E_{NES}$, where the small parameter $\varepsilon$ scales the mass of the NES and the stiffness. Physically, the appearance of the small parameter $\varepsilon$ implies that energy transfer occurs between the primary system and the NES on different time scales.

## 5.2. Energy transfer in the stationary oscillations

For the stationary oscillations, applying the Hilbert-Huang transform to $x_1$ and $x_2$ corresponding to the displacements for the primary system and the NES respectively, we get the instantaneous frequencies, $\omega_{LO}$, $\omega_{NES}$, with the change of time $\tau/2\pi$. In subfigures (a) and (b) of Fig.11, the Hilbert spectrums, consisting of the instantaneous frequencies ($\omega_{LO}$, $\omega_{NES}$), time ($\tau/2\pi$), and the instantaneous energies that are represented by the color bar, indicate that $\omega_{LO}$ and $\omega_{NES}$ both are nearly equal to 1 throughout the entire time range and remain unchanged. This means no energy transfer in the stationary oscillations.

Furthermore, we get the Hilbert spectrums corresponding to $E_{LO}$ and $E_{NES}$, respectively. $\chi_{LO}$ and $\chi_{NES}$ represent the instantaneous frequencies of the energies of the primary system and the NES, respectively. Viewed from subfigures (a) and (b) of Fig.12., the instantaneous frequencies $\chi_{LO}$ and $\chi_{NES}$ are equal to zero, indicating there is no energy transfer.

As we obtained in Section 4.2, in this case, the stationary oscillations are represented as a fixed point captured by the invariant manifold projected on the modulus plane ($N_{10}$, $N_{20}$). Thus, for this type of steady-state oscillations in the system, the amplitudes of the primary system response and the NES response are always unchanged except for an initial effluxion of time, under the external forcing. The theoretical results are confirmed by numerical simulations.

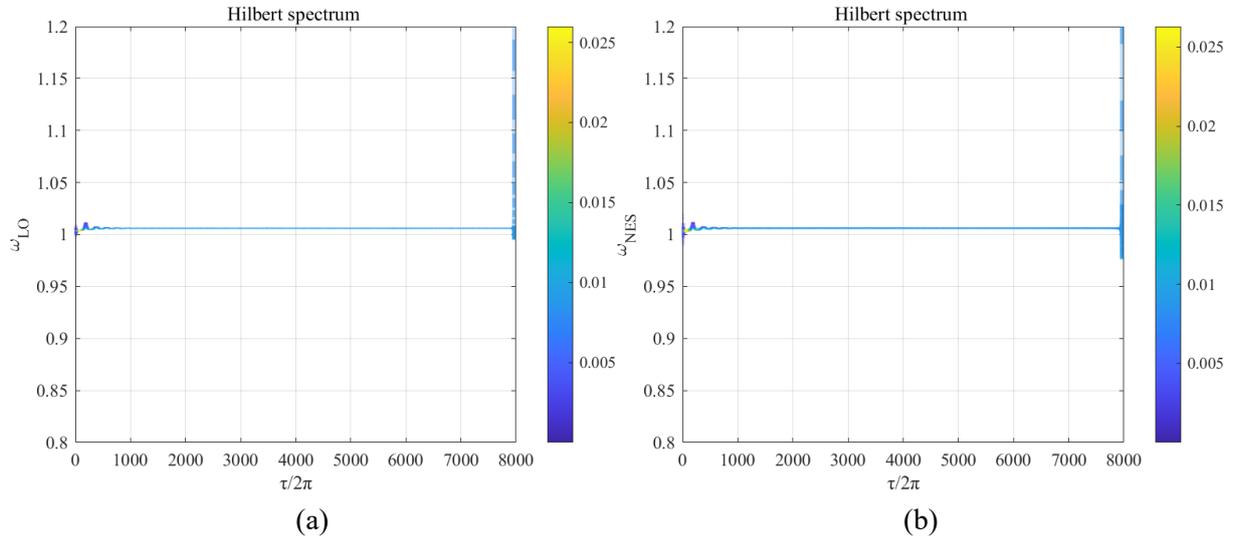

Fig.11. Hilbert-Huang transform for system response by fixing system parameter at $\varepsilon=0.001$, $k=0.5$, $k_2=0.5$, $\lambda=0.1$, $\lambda_1=0.1$, $\lambda_2=0.1$, $\sigma=6$ and $A=1$. (a) Hibert spectrum for the displacement $x_1$ of the primary system; (b) Hibert spectrum for the displacement $x_2$ of the NES.

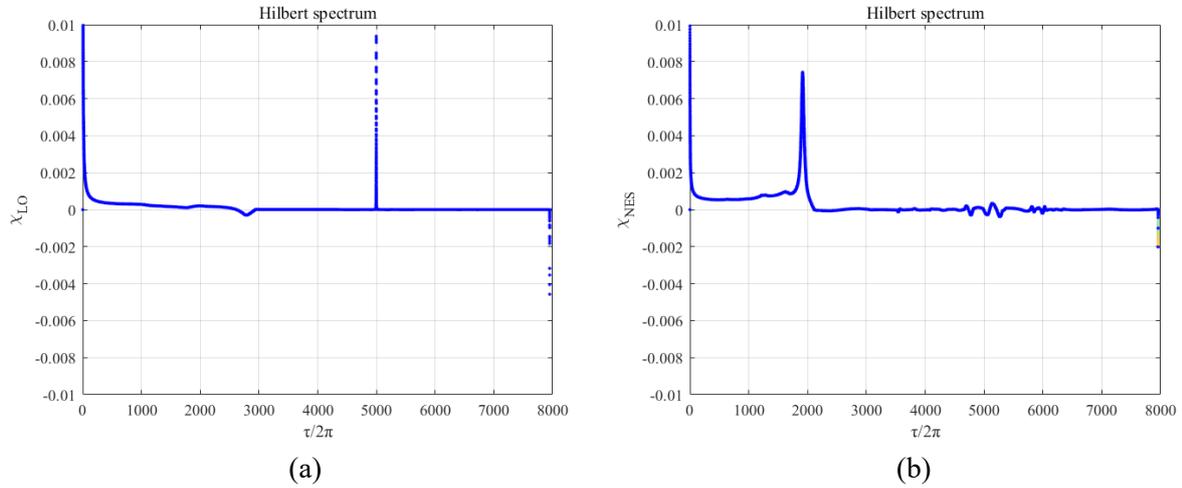

Fig.12. Hilbert-Huang transform for system energy by fixing system parameter at $\varepsilon=0.001$, $k=0.5$, $k_2=0.5$, $\lambda=0.1$, $\lambda_1=0.1$, $\lambda_2=0.1$, $\sigma=6$ and $A=1$. (a) Hibert spectrum for the energy $E_{LO}$; (b) Hibert spectrum for the energy $E_{NES}$.

### 5.3. Energy transfer in the nonlinear beats

For the nonlinear beats, we get the instantaneous frequencies, $\omega_{LO}$, and $\omega_{NES}$, by using the Hilbert-Huang transform to the displacements $x_1$ and $x_2$. In subfigures (a) and (b) of Fig.13., the Hilbert spectrums show tiny fluctuations, implying the occurrence of weak energy transfer in the nonlinear beats.

Moreover, the Hilbert spectrums corresponding to $E_{LO}$ and $E_{NES}$ show that energy transfer occurs between the primary system and the NES. In subfigures (a) and (b) of Fig.14., we can notice that the instantaneous frequencies $\chi_{LO}$ and $\chi_{NES}$ are changing over time. It is worth noting that the scale of this change is of the $O(\varepsilon)$ order, which indicates the occurrence of the $O(\varepsilon)/O(\varepsilon)$ type energy transfer between the primary system and the NES.

In Section 4.3, the trajectories of the amplitudes of the system responses are predicted by an interval of the invariant manifold. Thus, the two amplitudes both perform a slowly-varying change, leading to the evolutions of response amplitudes vibrating at the slow frequency $\omega_S$. In this case, the Hilbert spectrums show that the instantaneous frequencies of energies of the primary system and the NES both are $O(\varepsilon)$ order. This matches the theoretical prediction.

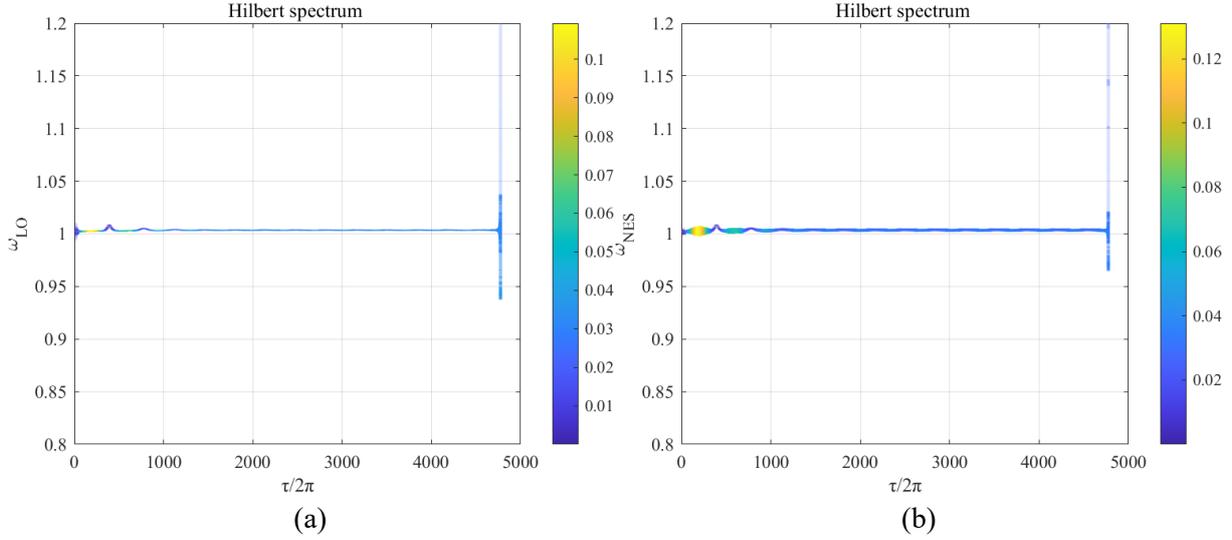

Fig.13. Hilbert-Huang transform for system response by fixing system parameter at $\varepsilon=0.001$, $k=0.5$, $k_2=0.5$, $\lambda=0.1$, $\lambda_1=0.1$, $\lambda_2=0.1$, $\sigma=3.5$ and $A=1$. (a) Hibert spectrum for the displacement $x_1$ of the primary system; (b) Hibert spectrum for the displacement $x_2$ of the NES.

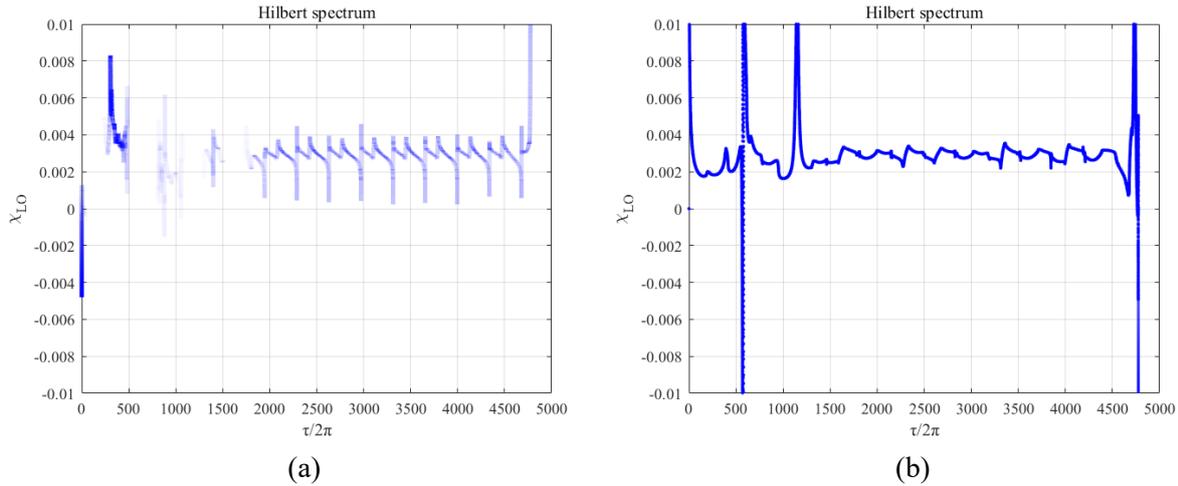

Fig.14. Hilbert-Huang transform for system energy by fixing system parameter at $\varepsilon=0.001$, $k=0.5$, $k_2=0.5$, $\lambda=0.1$, $\lambda_1=0.1$, $\lambda_2=0.1$, $\sigma=3.5$ and $A=1$. (a) Hibert spectrum for the energy $E_{LO}$; (b) Hibert spectrum for the energy $E_{NES}$.

### 5.4. Energy transfer in the relaxation oscillations

For the relaxation oscillations, it is necessary to emphasize that only the NES response is expressed as the relaxation oscillation, while the primary system response is expressed as the nonlinear beat. This means the occurrence of energy transfers between the primary system and the NES on different time scales. Applying the Hilbert-Huang transform to the displacements, $x_1$, $x_2$, of the system, we get the Hilbert spectrums, depicted in Fig.15. In subfigure (a) of Fig.15., the instantaneous frequency $\omega_{LO}$ is approximately equal to 1 over time, but the instantaneous energy $E_{LO}$ has changed since its color has changed over time. In subfigure (b) of Fig.15., one can find the instantaneous frequency $\omega_{NES}$ exhibits varying periodically and makes fast oscillations.

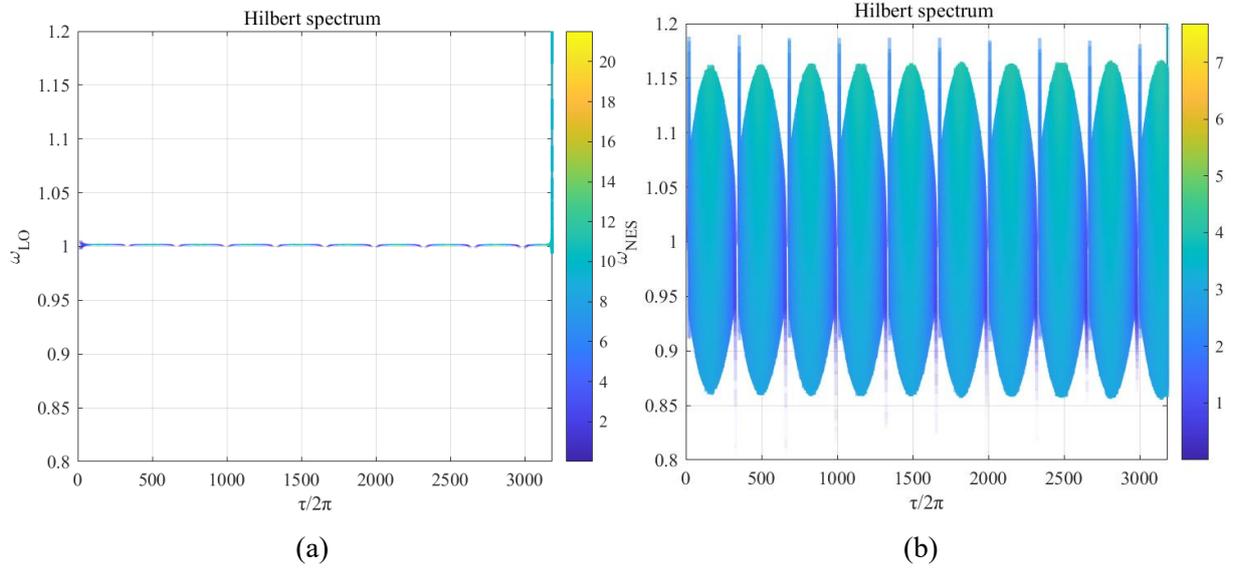

Fig.15. Hilbert-Huang transform for system response by fixing system parameter at $\varepsilon=0.001$, $k=0.5$, $k_2=0.5$, $\lambda=0.1$, $\lambda_1=0.1$, $\lambda_2=0.1$, $\sigma=3.14$ and $A=10$. (a) Hibert spectrum for the displacement $x_1$ of the primary system; (b) Hibert spectrum for the displacement $x_2$ of the NES.

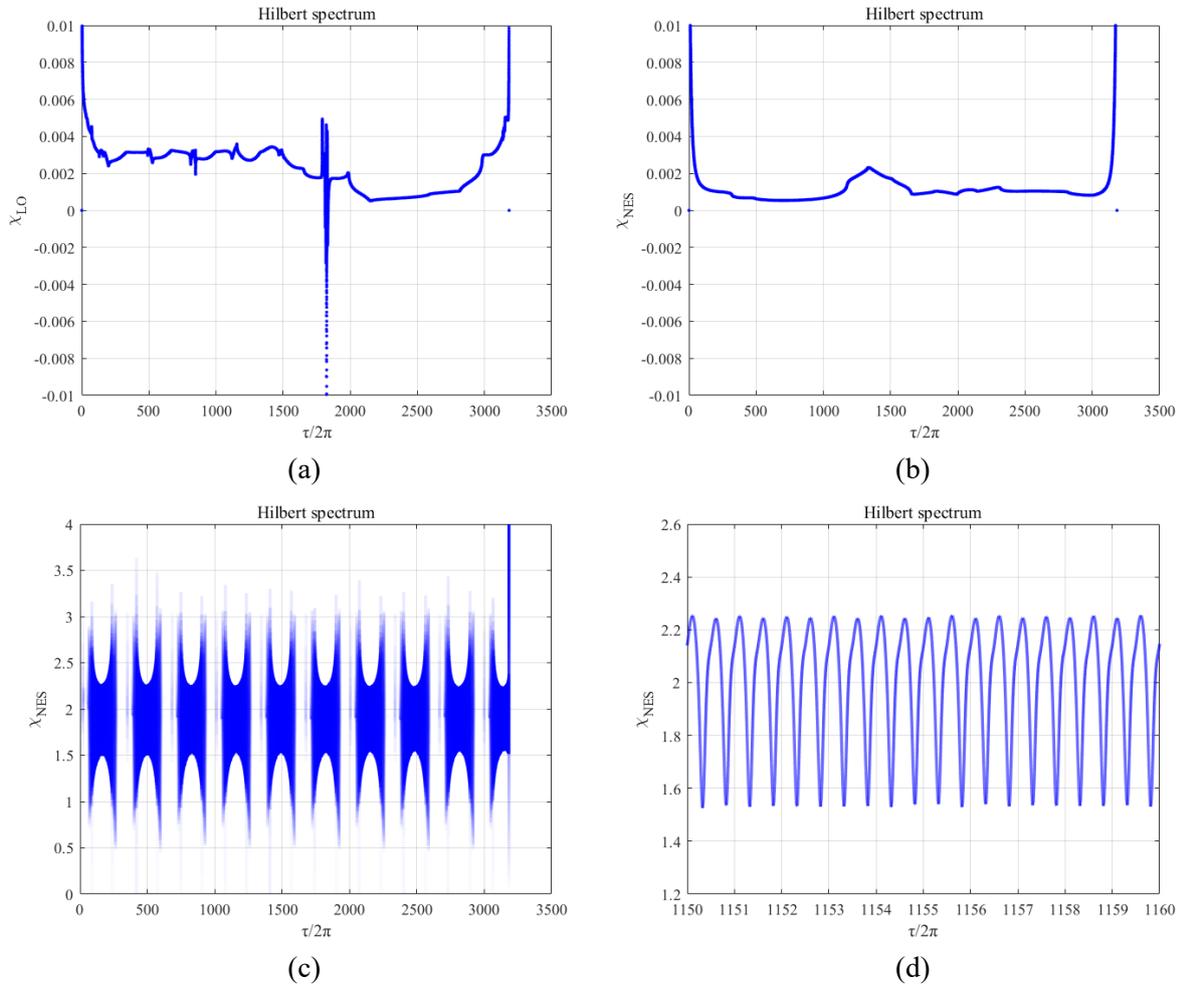

Fig.16. Hilbert-Huang transform for system energy by fixing system parameter at $\varepsilon=0.001$, $k=0.5$, $k_2=0.5$, $\lambda=0.1$, $\lambda_1=0.1$, $\lambda_2=0.1$, $\sigma=3.14$ and $A=10$. (a) Hibert spectrum for the energy $E_{LO}$; (b) Hibert spectrum for the energy $E_{NES}$ on the slow time scale; (c) Hibert spectrum for the energy $E_{NES}$ on the fast time scale; (d) Close up.

Further, taking the Hilbert-Huang transform to the energies, $E_{LO}$, $E_{NES}$, and the corresponding Hilbert spectrums can be obtained. For the instantaneous frequency $\chi_{LO}$ of $E_{LO}$, there are only the values occurring in the $O(\varepsilon)$ order, seeing in subfigure (a) of Fig.16. However, for the instantaneous frequency $\chi_{NES}$ of $E_{NES}$, it has not only the values in the $O(\varepsilon)$ order but also the fast oscillations occurring in the $O(1)$ order. The $O(\varepsilon)$ order and the $O(1)$ order for the instantaneous frequency $\chi_{NES}$ are depicted in subfigures (b) and (c) of Fig.16., respectively. And, in subfigure (d) of Fig.16., we can find the fast oscillations in the $O(1)$ order of $\chi_{NES}$. According to that, in the relaxation oscillations, the energy transfers between the primary system and the NES involve both the $O(\varepsilon)/O(\varepsilon)$ and the $O(\varepsilon)/O(1)$ types. Therefore, we realize that the energy transfers occurring in the relaxation oscillations are multiscale and intensive.

It is worth noting that the changes in the amplitude of NES response on the fast time $\tau_0$ and the slow time $\tau_1$ occur simultaneously. As we analyzed in Section 3.3, while the complex amplitude of the NES triggers the high-order frequency term $G_3$, it should simultaneously satisfy the condition of the invariant manifold $G_1=0$. The results of the analytical study of the type of energy transfers in the relaxation oscillations are in agreement with the results of numerical simulations.

# 6. Conclusions

This work has revealed the occurrence of multiscale energy transfers in the two-DOF mechanical system, composed of a primary system and a grounded NES. Using the complex coordinates and the multiple time scales method, the slow frequency and the fast frequency of the complex amplitudes have been obtained.

Furthermore, it has been shown that steady-state oscillations can be roughly divided into three types, namely stationary oscillations, nonlinear beats, and relaxation oscillations. The corresponding forms of the system are rewritten to match the properties of the three steady-state oscillations. For the stationary oscillations, there is no variation in the amplitudes of the system responses. For the nonlinear beats, there is a slow frequency in the amplitudes of the primary system and the NES. For the relaxation oscillations, the system responses contain not only the slow frequency but also the fast frequency. It is worth noting that the nonlinear beats and the relaxation oscillations are non-stationary oscillations, but both of them belong to steady-state oscillations since they both satisfy Eq.(48).

*Assumption 1* and *Assumption 2* in Section 3.1 have been confirmed by numerical results. The numerical results show that all branches of the invariant manifold contain slow frequency, and only the upper branch contains fast frequency triggered by the higher-order frequency term $G_3$. Finally, several types of energy transfer were obtained by HHT. It was shown that there is no energy transfer in the stationary oscillations, the $O(\varepsilon)/O(\varepsilon)$ type of energy transfer in the nonlinear beats, and both the $O(\varepsilon)/O(\varepsilon)$ and the $O(\varepsilon)/O(1)$ types of energy transfers in the relaxation oscillations. The theoretical results are confirmed by numerical simulations.

# CRediT authorship contribution statement

**Lan Huang:** Conceptualization, Methodology, Investigation, Writing – original draft.

**Xiaodong Yang:** Conceptualization, Supervision, Validation, Writing – review & editing.

# Declaration of Competing Interest

The authors declare that they have no known competing financial interests or personal relationships that could have appeared to influence the work reported in this paper.

# Data Availability

Data will be made available on reasonable request.

# Acknowledgment

This work was supported by the National Natural Science Foundation of China (No.12332001; No.11972050).